\documentclass[12pt]{article}
\usepackage{amsfonts}
\usepackage{eurosym}
\usepackage{graphicx}
\usepackage{booktabs}
\usepackage{float}
\usepackage{subfig}
\usepackage{amssymb}
\usepackage{comment}
\usepackage{amsmath}
\usepackage{booktabs}
\usepackage{multirow}
\usepackage{enumerate}
\usepackage{color}
\usepackage[authoryear]{natbib}
\usepackage{hyperref}

\hypersetup{colorlinks=true, linkcolor=red, citecolor=blue}

\newtheorem{lemma}{{\bf \sc Lemma}}

\newtheorem{proposition}{{\bf \sc Proposition}}

\def\Re{I\!\!R}
\def\eproof{\hbox{\hskip3pt\vrule width4pt height8pt depth1.5pt}}
\def\E{{\mathrm{E}}}
\def\Var{{\mathrm{Var}}}
\begin{document}

\title{The Friendship Paradox and Systematic Biases in Perceptions and Social Norms}
\author{Matthew O. Jackson \thanks{%
Department of
Economics, Stanford University, Stanford, California 94305-6072 USA,
external faculty member at the Santa Fe Institute, and a fellow of CIFAR.
Email:  and jacksonm@stanford.edu.  I gratefully
acknowledge financial support under ARO MURI Award No. W911NF-12-1-0509 and NSF grant SES-1629446.
I thank Ozan Candogan, Itay Fainmesser, Emir Kamenica, Sharon Shiao, Adam Szeidl, Xu Tan, Yiqing Xing, Yves Zenou, and two referees, as well as the participants of WISE 2016, the 2017 Social Networks and Information Conference, and Network Science in Economics 2017, for very helpful comments and suggestions.
}}
\date{Draft: November 2017}
\maketitle

\begin{abstract}

The ``friendship paradox'' (\cite{feld1991}) refers to the fact that, on average, people have strictly fewer friends than their friends have.  I show that this over-sampling of the most popular people amplifies behaviors that involve complementarities.  People with more friends experience greater interactive effects and hence engage more in socially influenced activities.  Given the friendship paradox, people then perceive more engagement when sampling their friends than exists in the overall population.  Given the complementarities, this feeds back to amplify average engagement.   In addition, people with the greatest innate benefits from a behavior also tend to be the ones who choose to interact the most, leading to further feedback and amplification.  These results are consistent with studies finding that people consistently overestimate peer consumption of alcohol, cigarettes, and drugs; and, can help explain problems with adolescent abuse of drugs and binge-drinking, as well as other behaviors.  I also discuss how these results change in cases of strategic substitutes, where individuals overestimate free-riding by peers.

\textsc{JEL Classification Codes:} D85, D13, L14, O12, Z13

\textsc{Keywords:} Social Networks, Social Norms, Friendship Paradox, Networks, Games on Networks, Complementarities, Peer Effects, Public Goods, Network Formation, Social Norm Marketing
\end{abstract}

\thispagestyle{empty}

\setcounter{page}{0} 

\section{Introduction}

Social norms are governed by our perceptions of others, which are heavily determined by those around us.\footnote{For a broader discussion of norms and perceptions, see \citet{hanh2016}.}
However, our friends are not a random selection from the population: {\sl even on average} we are biased in the samples with whom we
interact.  This  can systematically distort our beliefs and affect our behaviors, ranging from
consumption of cigarettes, alcohol, and drugs by adolescents to our propensity to donate to charities.

The distortion stems from the  ``friendship paradox''  that was pointed out by the sociologist Scott Feld in \citeyear{feld1991}.
Feld observed that people's friends have more friends than people do, on average.
That is, the average number of friends that a typical person's friend has is greater than the average number of friends that people have in the population.
This follows from the fact that a person with many friends is observed by more people than someone who has few friends, and so people's samples of friends
are weighted by friends' popularity rather than by their
proportions in the population.

The extent of the friendship paradox varies by setting, but is present in {\sl every} network in which there is at least one friendship between two people who have different degrees (numbers of friends), as proven in Lemma \ref{paradox2} below.
As an example, I show that in a rural Indian village, friends have on average more than 40 percent more friends than the average villager.
The friendship paradox is greatly magnified by social media: a study of Twitter behavior by \citet*{hodaskl2013}
found that more than 98 percent of users had fewer followers than the people whom they followed: typically a user's ``friends'' had 1000 percent more followers, or more, than the user.
Given the increased use of social media, especially by adolescents, the potential for biased perceptions in favor of a tiny proportion of the most popular users becomes overwhelming.

Since many risky and addictive behaviors are peer influenced, and driven by people's perceptions of what is normal or acceptable behavior,
it is important to understand how perceptions are formed and how they may impact behaviors.
The impact of the friendship paradox on such behaviors can be seen a series of studies that find that students tend to over-estimate the frequency with which their peers smoke and consume alcohol and drugs, and often by
substantial margins.  For instance, a study covering one hundred U.S. college campuses by \citet*{perkinsmlcp1999}
found that students systematically over-estimate consumption of eleven different substances, including cigarettes, alcohol, marijuana, and a variety of other drugs.\footnote{As this is survey-based data, it could be that students under-report their own consumption but not their perception of others.
As discussed in \cite{perkinsmlcp1999}, care was taken to make the surveys anonymous, and comparisons to other data offer some reassurance that the distorted reporting is unlikely to
account for the full discrepency.  Nonetheless, the findings should be interpreted cautiously.  }
An idea of the magnitude of the effects can be seen in alcohol consumption by comparing students self-reported drinking behavior - how many drinks they had the last time they partied or socialized - to their perceptions of how many the typical student at their school the last time she or he partied or socialized.  The median student (out of the
more than 72000 students on the 130 colleges in the study conducted from 2000 to 2003) reported consuming 4 drinks and a quarter of the students reported consuming 5 drinks or more.
However, even though these numbers are more than double the CDC's (Center for Disease Control's) recommended limits,\footnote{https://www.cdc.gov/alcohol/fact-sheets/moderate-drinking.htm
 accessed May 6, 2017.} more than 70 percent of the students still managed to over-estimate the alcohol consumption of the typical student at their own school by at least 1 drink and 39 percent overestimated the norm by 3 or more drinks (see Table 2 in \citet*{perkinsh2005}).\footnote{This is but one application out of many in which perceived norms influence people's behaviors.  Others with negative externalities and complementarities include things like corrupt behavior, tax evasion,  and speeding.
 Of course, the model also applies to settings with positive externalities and complementarities (as discussed when analyzing welfare), and includes things like study habits, human capital accumulation, charitable behavior, among others.}

To explain these sorts of misperceptions, we don't have to dig deeply into the psychology of the students.
When students are attending parties or social events, they are interacting disproportionately with the people who attend the most parties - so students' perceptions of alcohol consumption end up over-representing the people who attend more parties and under-representing those who attend infrequently.

This bias in observation due to the friendship paradox would not have any impact, however, unless those who have more connections end up behaving
systematically differently from those with fewer connections.   For instance, in order for
 the friendship paradox to matter, it has to be that more popular students are more likely to smoke or consume alcohol in order to bias
students' estimates upwards.
Indeed,  there is empirical evidence for this in the context of student consumption of drugs and alcohol.
For example, \citet*{valenteuj2005}, in a study of middle school students, found that each additional friendship accounted for a
5 percent increase in the probability that a student smoked.
\citet*{tuckermdzgs2013} found similar numbers for alcohol,
finding that being named as a friend by five additional others
accounted for a 30 percent increase in the likelihood that a middle school student had tried alcohol.

The point of this paper is to explore the impact of the friendship paradox on behavior and explain why we should expect
more connected individuals to behave systematically differently from less connected agents,  and then how this feeds back to inflate overall behavior.
As I establish here, there are two basic forces at work.   One is that people who have the most connections are most exposed to interactions with others, and thus in any setting of strategic complements (or substitutes), their behaviors are most heavily influenced.
The second is that if people differ in their tastes for a given activity, then  it is the people who benefit most from that activity who choose to have the most connections.  So, if we endogenize the network, individuals with a  greater marginal payoff from a given activity end up having the most influence on others.
This further amplifies the effect, increasing the disparity of actions between high-degree and low-degree individuals.
Combined, these forces lead people's most popular friends to engage the most in a behavior, and via feedback through the complementarities to bias the overall
behavior in the society.\footnote{Throughout the paper, I refer to `more' engagement or `higher' levels of activity as the result of the complementarity.  As usual in games with complementarities, this is just a sign convention as, for instance, spending less time studying is equivalent to spending more time not-studying.}
For instance, returning to the example above, since consuming alcohol by teenagers is in part (or largely) a social activity, the people who spend more time socializing with others would have more reason to consume alcohol at an early age, and would also tend be those who have a greater base proclivity to consume alcohol at an early age.
So, students who are more often seen as friends by others being more likely to consume alcohol leads to biased samples and biased perceptions, consistent with the data, and feeds back to produce high levels of activity overall.

The proofs are intuitive and direct, which should not detract from the results.   The contribution of the paper is less in developing a new methodology, than in presenting a clear and simple framework in which we can see how the friendship paradox leads to systematic
biases in any setting with complementarities.
The point is still important to explore and understand given its wide-ranging implications.

Before describing the formal model, let me begin with some background on the friendship paradox and a simple illustration of how it can bias behaviors in the context of an example.

\subsection{The Friendship Paradox}

\begin{figure}[h!]
\begin{center}
\includegraphics[height=2in]{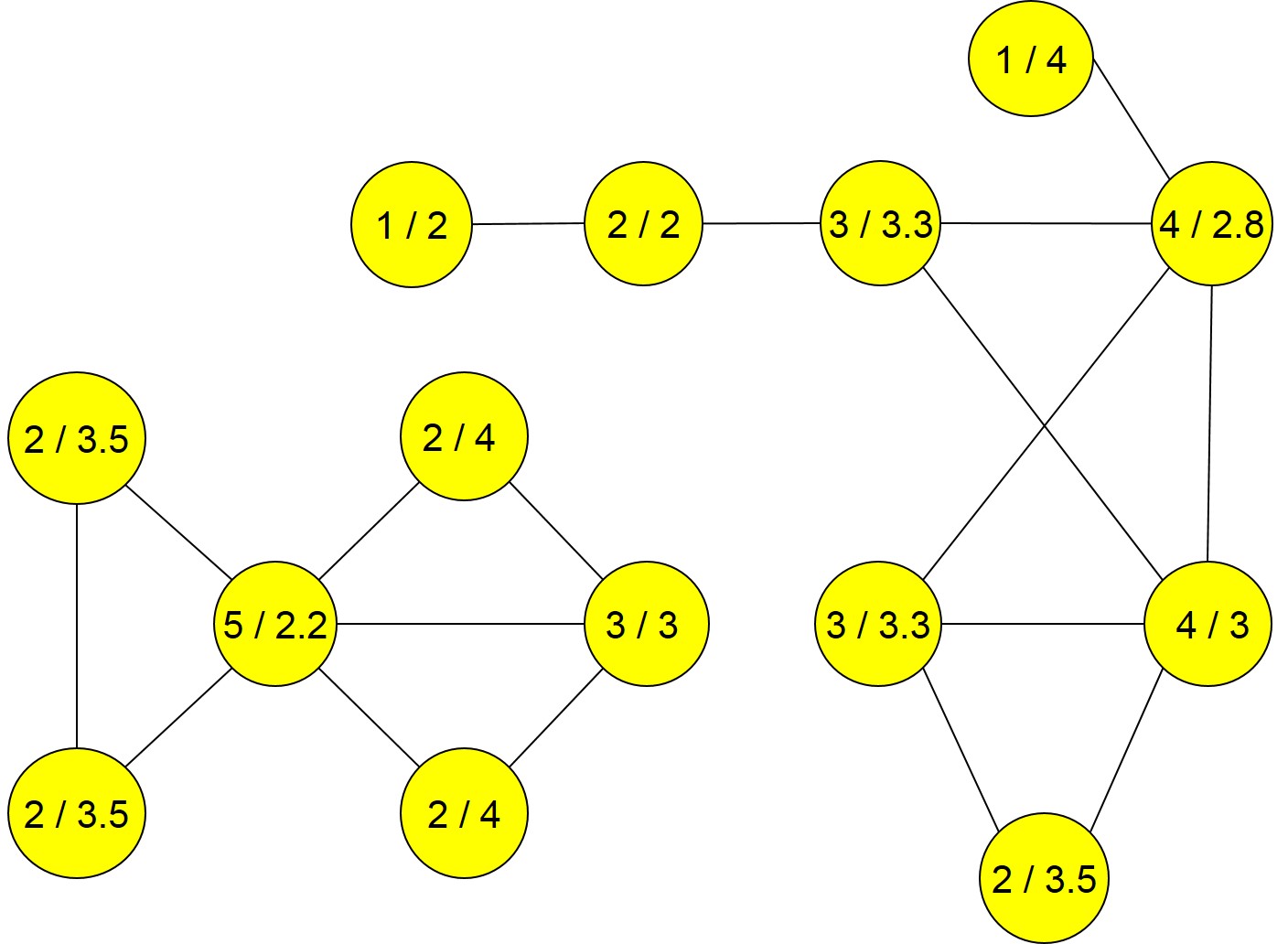}
\caption{\label{figurefriendship-paradox1}
\small{Data from James Coleman's \citeyearpar{coleman1961} study of high school friendships.
Nodes are girls and links are mutual friendships.
The first number listed for each girl is how many friends the girl has and the second number is the average number of friends that the girl's friends have.
For instance, the girl in the lower left-hand corner has 2 friends, and those friends have 2 and 5 friends, for an average of 3.5.
}}
\end{center}
\end{figure}

Let us begin with a quick look at
the data set from  \citet{coleman1961} that was originally cited by \cite{feld1991}.
A portion of Coleman's data is pictured in Figure \ref{figurefriendship-paradox1}.\footnote{These are just two components of the network.  The larger network not pictured here  exhibits the same phenomenon:  146 girls have friends (defined mutually), and of those, 80 have fewer friends than
their friends on average, while 25 have the same number as their friends, and 41 have more friends than their friends. 
}
There are fourteen girls pictured.   For nine of the girls, their friends have on average more friends than they do.  Two girls have the same number of friends as their friends do on average, while only three of the girls are more popular than their friends on average.  On average the girls have 2.6 friends, while on average their friends have 3.2 friends.

\begin{figure}[h!]

\centering
\subfloat[Histogram:  Blue = Own Degree,  \, \, \, \, \, \, \, \, \, \ \ \ \ \ \ \ \ \ \ \ \ \
Red = Avg. Neighbors' Degree. ]{
\label{fig:h1}
\includegraphics[width=0.45\textwidth]{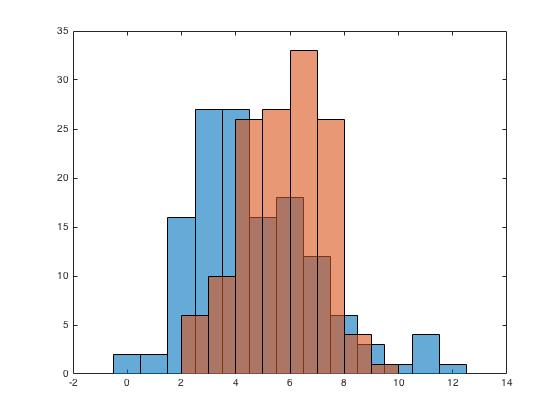}
}
\subfloat[Histogram: Ratio of Average Neighbors' Degree over Own Degree.]{
\label{fig:h2}
\includegraphics[width=0.45\textwidth]{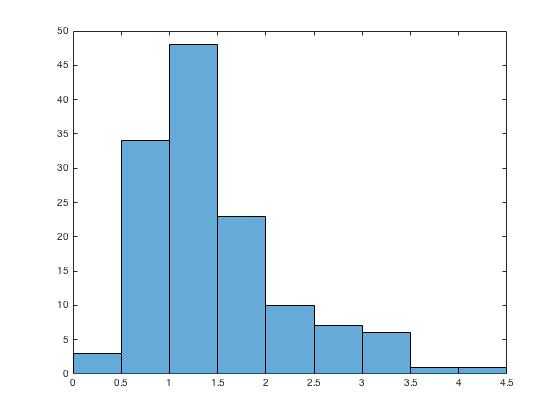}
}
\caption{\label{figurevillage26}
 The Friendship Paradox illustrated in a network of 135 households from a typical rural Indian Village in the study of \citet*{banerjeecdj2013}.  Nodes are households and
links are other households with whom the household exchanges favors (borrowing/lending kerosene and rice).
Panel (a) compares the distribution own to average neighbors' degrees in the network. 
Panel (b) shows the distribution of the ratio of the average of neighbors' degrees compared to own degree, with an average ratio of 1.43.
}
\end{figure}

To see the friendship paradox in more detail and in a larger network, consider a network of connections between households from a rural Indian village.
The full distribution of degrees and the distribution of degrees of neighbors is given in Figure \ref{figurevillage26}, and we see that friends' degrees are more than forty percent larger than the average degree in the society.

The friendship paradox is easy to understand.  The most popular people appear on many other people's friendship lists, while people with very few friends appear on relatively few people's lists.
The following lemma provides a general statement of the friendship paradox, showing that it holds in all networks.  One can find a variation of this lemma in \citet*{justcl2015}.

A finite set $N=\{1,\ldots, n\}$ of agents, with generic indices $i,j$, are members of an undirected (having consensual friendships)
network $g\in \{0,1\}^{n\times n}$, so that $g$ is symmetric and has 0's on the diagonals.\footnote{For omitted definitions, see \citet{jackson2008}. The paradox extends to directed networks when one considers the average in-degree of friends.}
Agent $i\in N$ has $d_i(g) = \sum_j g_{ij} $  links (friendships) in the network.

\newpage

\begin{lemma}
\label{paradox2} [The Friendship Paradox - General Networks]

For any network, the average degree of neighbors is at least as high as the average degree and the inequality is strict if and only if at least two linked agents have different degrees.
That is,  $\frac{1}{n}\sum_{i: d_i(g)>0} \frac{\sum_{j: g_{ij}>0} d_j(g)}{d_i(g)}  \geq \frac{\sum_{i} d_i(g)}{n} $, with strict inequality  if (and only if) at least two linked agents have different degrees.
\end{lemma}

The proof is straightforward and for completeness appears in the appendix.  A stronger characterization of the magnitude of the friendship paradox appears in Lemma \ref{friendshipparadox} below, as it can be derived once we give more structure to the set of networks considered.  Other variations on the friendship paradox and bounds on its magnitude in specific models can be found in \citet[Section 4.2.1]{jackson2008}, \citet{lattanzis2015}, \citet{caor2016}.

This paradox, although easily understood, has wide-ranging implications, as we shall see.

\subsection{An Example of the Impact of the Friendship Paradox}\label{exampleimpact}

To see the implications of the friendship paradox most starkly, let us consider a simple example.

A society of agents are influenced by their friends.\footnote{%
To see similar examples illustrating biased estimation of opinions, see \citet*{lermanyw2015}.  On can also find examples in popular blogs (e.g., see Kevin Schaul's {\sl Washington Post} blog from Oct. 9, 2015
``A quick puzzle to tell whether you know what people are thinking'').}
The agents choose one of two actions, either solid or plaid.
They each have a slight preference for solid or plaid and in the first period they follow those preferences.   However, agents are conformists and prefer to match the majority of others,
and only follow their own preference if there were equal numbers of others in each style.
They start with the choices in the upper left-hand figure, with only four people preferring solid and eight preferring plaid.   If they could all see
the whole group and best replied to that, then they would all choose plaid in the next period.
However, instead
agents actually see and react to their neighbors in the network.

\begin{figure}[h!]
\centering
{
\label{fig:image_3a}
\includegraphics[width=0.45\textwidth]{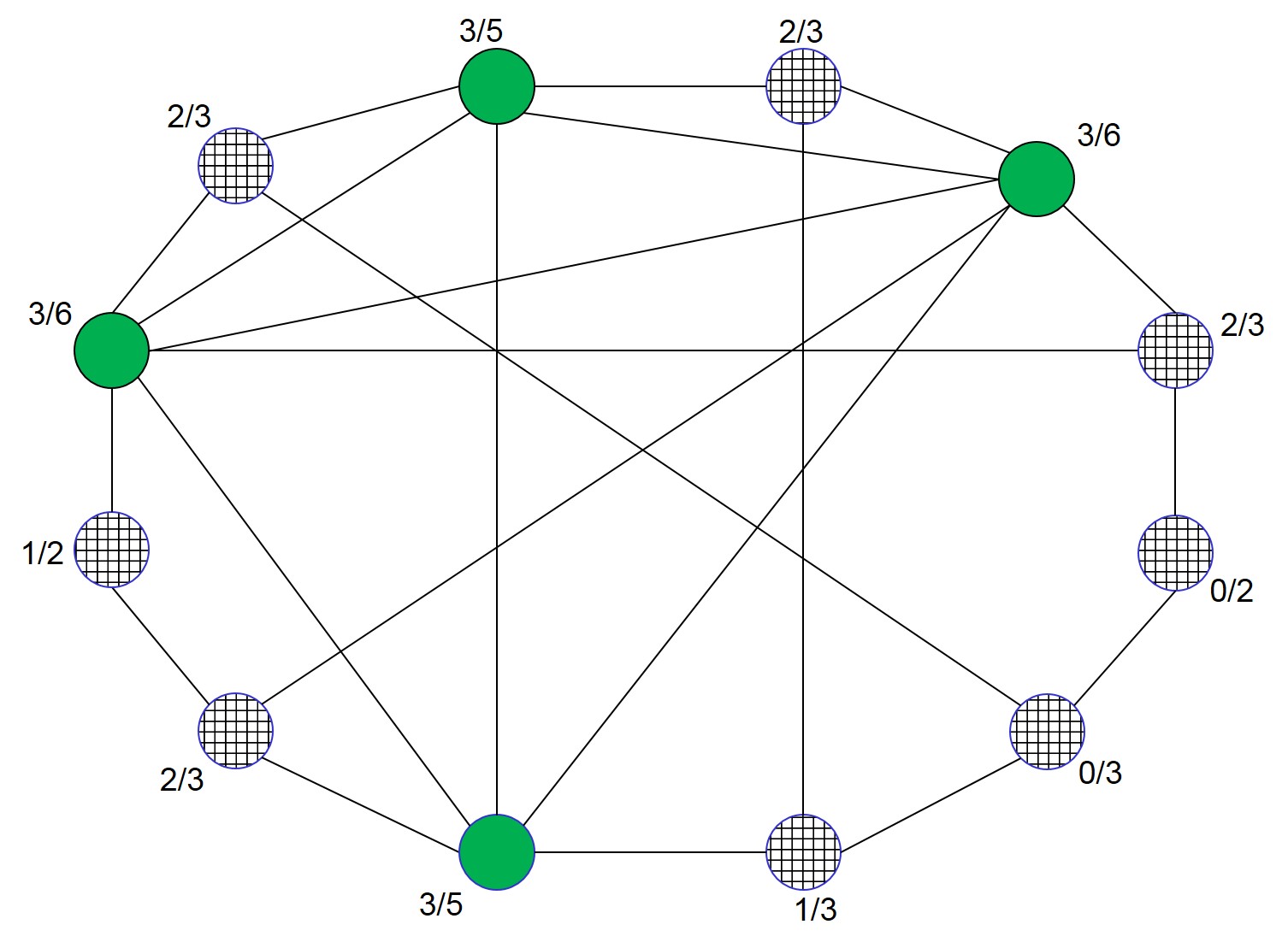}
}
\caption{\label{figurefashion1} The Friendship paradox at work.   The four most connected agents have a base preference for solid and the eight others prefer plaid.
The fractions next to the agents are their perceptions of the preferences are for solids over
plaids, based on what they see among their friends in the first period.  Most of them perceive a majority preference for solid, with only the few agents in
the lower right perceiving a majority for plaids. }
\end{figure}

\begin{figure}[h!]
\centering
\subfloat[Day 2, Four plaids switch to match the popular agents.]{
\label{fig:image_3}
\includegraphics[width=0.45\textwidth]{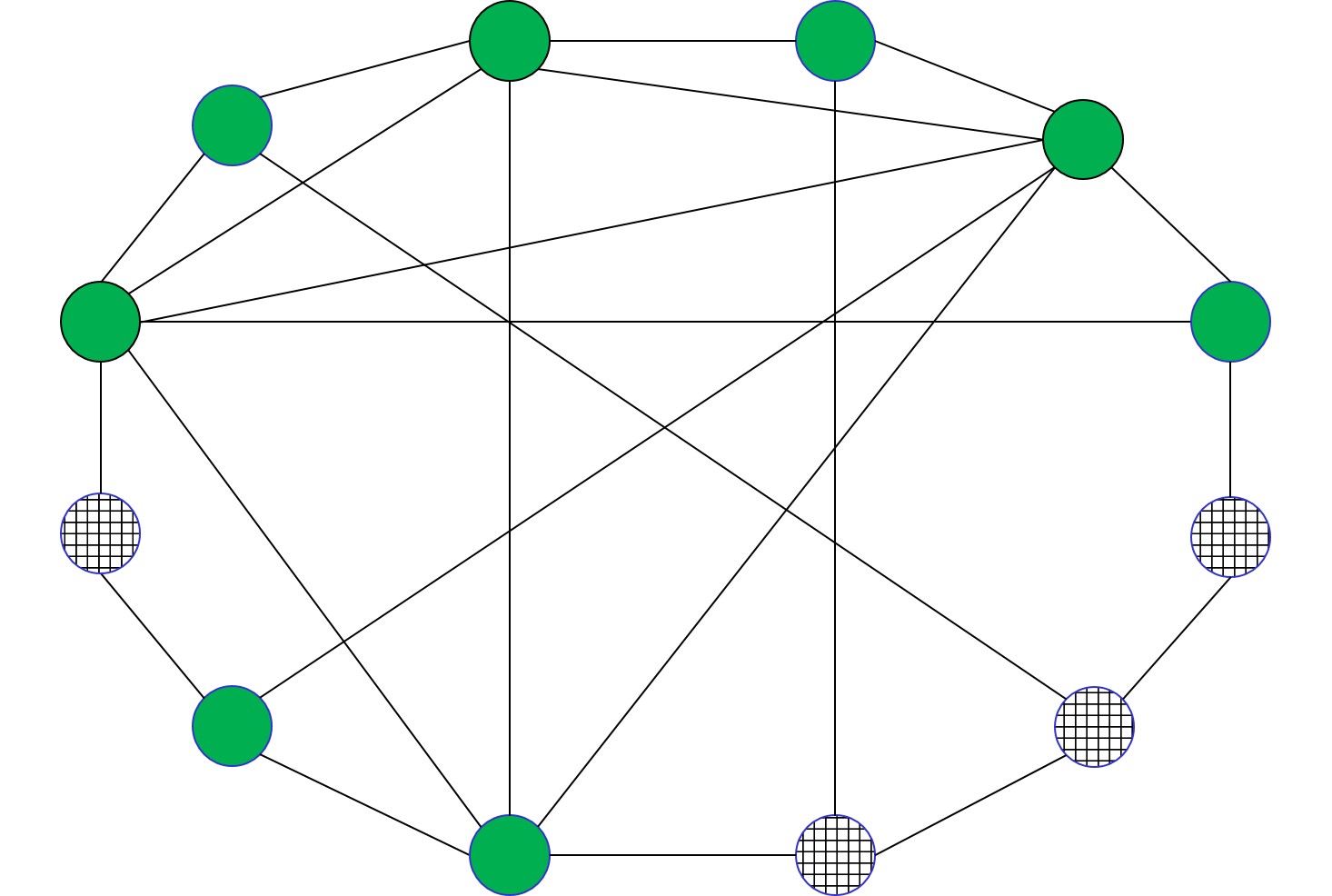}
}
\subfloat[Day 3, More switch.]{
\label{fig:image_4}
\includegraphics[width=0.45\textwidth]{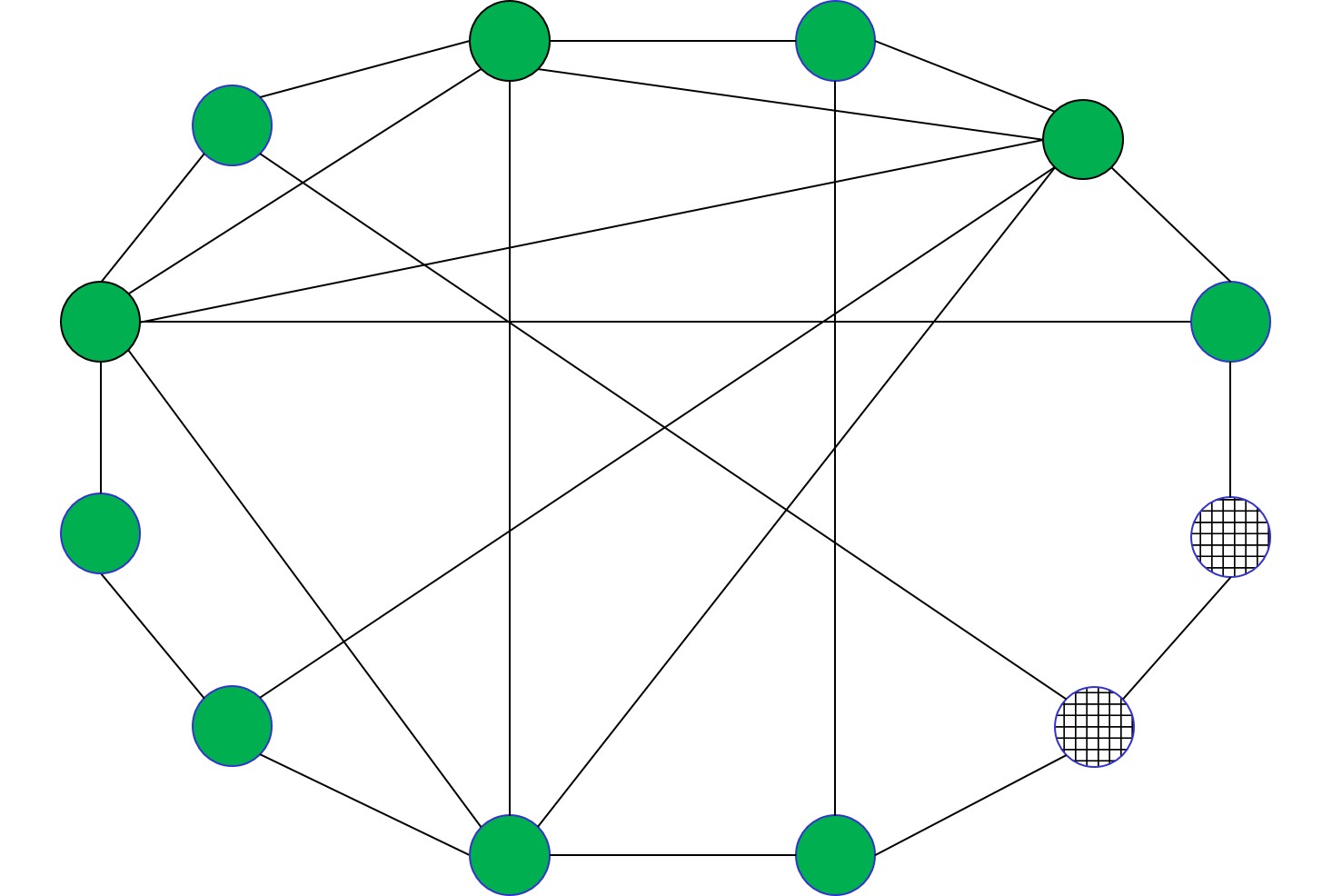}
}

\centering
\subfloat[Day 4, The changes to solids continue.]{
\label{fig:image_5}
\includegraphics[width=0.45\textwidth]{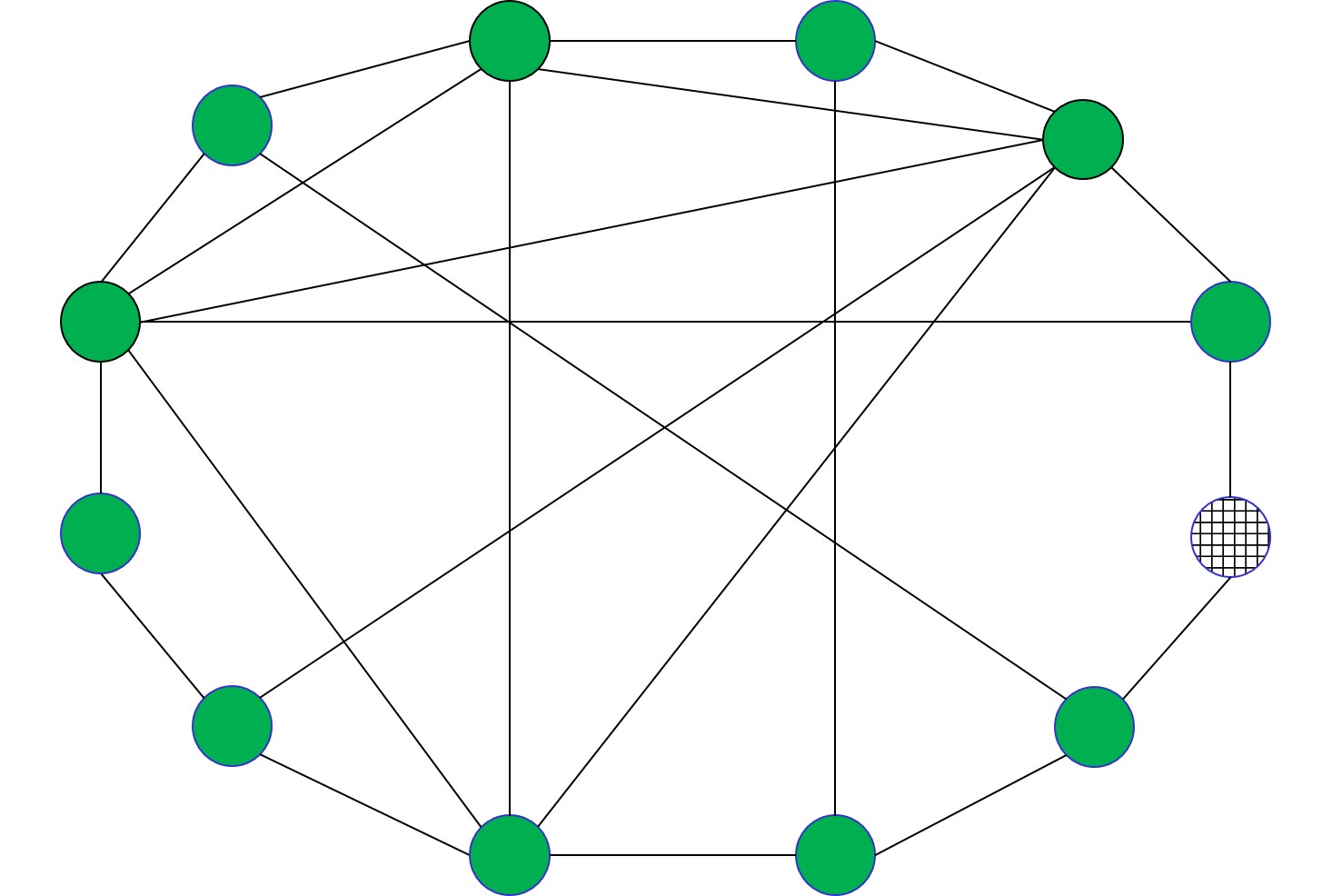}
}
\subfloat[Day 5, All agents conform to solids.]{
\label{fig:image_6}
\includegraphics[width=0.45\textwidth]{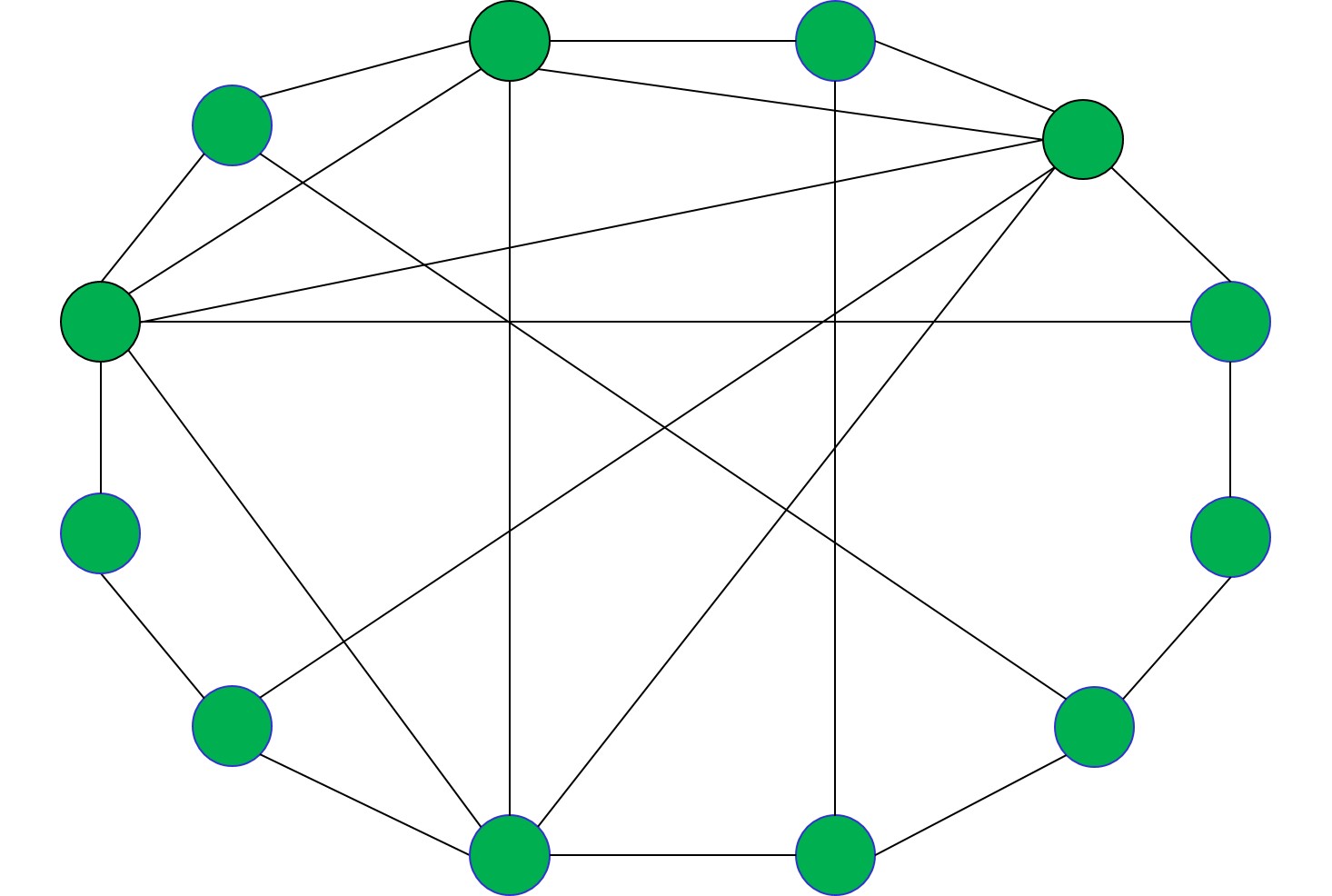}
}
\caption{\label{figurefashion2} Best reply dynamics:  agents wish to match the behaviors of their neighbors and use their own preference to break ties.   The most popular are all friends with each other and all stay with solid.  Popular students are over-represented in other agents' neighborhoods and perceptions, and lead a cascade to solid.
}
\end{figure}

We start with the four most popular agents preferring solid, as pictured in Figure \ref{figurefashion1}.
The remaining figures show what happens each following period under a best-reply dynamic, in each period agents best respond to the choices that they see among their neighbors in the previous period.
So, Figure \ref{fig:image_3} has the best responses to Figure \ref{figurefashion1}.
The popular agents all see each other and some others, but a majority of whom they see are solid and so they stay with solid.
Some other agents react to the popular agents and switch to solid.
Iterating on this in Figures \ref{fig:image_4} to \ref{fig:image_6}, solid cascades and becomes the unanimous choice.

We see the friendship paradox's role by re-examining Figure \ref{figurefashion1}, which shows agents' perceptions of the fraction preferring plaid based on their observations of their friends.  Three quarters
of them see at least half preferring solid even though the actual population fraction is only one third.

The effect in this example is consistent with a set of experiments by
\citet*{kearnsjtw2009}.
They set up a laboratory version of a committee or political party having to agree on a candidate, either red or blue.
Groups of 36 subjects had to coordinate unanimously on a candidate in order to get paid.   Like our solid-plaid example above, the subjects were connected in a network.
They were at computer screens and could each toggle back and forth between red or blue at any instant.
They could also see which color their friends in the network were supporting at any time; but they could not see the choices of any other subjects beyond their friends.
Their objective was to reach a consensus within 60 seconds.
If all 36 subjects ever managed to reach the same color at some instant,
then the experiment ended with that being the consensus.
If the subjects came to a consensus, unanimously supporting the same candidate, then they won a monetary payment.  If they did not reach a consensus then they did not receive the payment.
Again, similar to our solid-plaid example, the subjects had preferences over the candidates.
Some subjects received a higher monetary payment if
the red candidate was the consensus and others got paid more if the blue candidate was the consensus.
For instance, in some treatments, a red-supporting subject got paid fifty cents if the group unanimously supported the blue candidate and a dollar fifty if the group unanimously supported the red candidate, and nothing if the group failed to reach unanimity.
Thus, subjects preferred to have a consensus on their preferred candidate, but would rather reach a consensus on the other candidate than to fail to reach a consensus.

There were 27 runs of the experiment in which the network was set up in a manner similar to our solid-plaid example:  only a  small minority of the subjects were supporters of one color and vast majority of subjects were supporters of the other color.\footnote{The precise mix varied across iterations of the experiments:  for
instance with 6 subjects preferring red and 30 preferring blue; or 9 preferring red and 27 preferring blue, or 14 preferring red and 22 preferring blue. Which color was the minority was randomized across experiments so that some bias in people's intrinsic preferences over colors did not bias the results.}
The key was that the small minority of subjects preferring red were the `popular' nodes - having many more friends in the network.

A consensus was reached in 24 out of the 27 iterations of the experiment.  More importantly, {\sl every} one of the successful groups reached a consensus that was the minority group of `popular' subjects' preferred candidate, even though the majority preferred the other candidate.  So, consistently,
even when only six subjects preferred one color, and thirty preferred the other, the group still settled on the preferred color of the small group of
the most popular subjects.\footnote{There is also evidence that the friendship paradox helped with coordination.
\citet{kearnsjtw2009} ran some other variations of the experiment in which  the networks were instead structured to be more evenly balanced with more equal numbers of supporters of red and blue, and in which the subjects had similar numbers of connections - so without a set of popular subjects.  In those versions of
the experiment the coordination was significantly less likely to occur:  only 11 out of 27 groups managed to coordinate when the number of connections was fairly evenly balanced among subjects.}

\subsection{The Contribution of the Current Paper}

The experiments and example described above show how the friendship paradox {\sl can} matter.
However, the example and experiment have starting behavior that is correlated with degree.
What is missing is an understanding of why higher degree individuals' actions should exhibit any systematic pattern that differs from others, and how this feeds back to the society.
If we had begun with higher degree individuals split evenly between solid and plaid in our example, or blue and red in the experiments, then there would have been no predictable bias in the outcome.

The contribution of this paper is to show why the friendship paradox matters by embedding it in settings in which agents' behaviors are influenced by their friends and
the overall level of activity of their friends.   This builds on a previous literature that has established comparative statics in games of strategic complements, including \citet*{jacksony2005,jacksony2007,ballestercz2006,galeottigjvy2010,bergemannm2013,bramoullek2014}.
Generally, in games with strategic complements or substitutes, higher degree individuals are exposed to more activity and are more affected.
This leads them
to engage systematically more in a behavior, which then feeds back to increase overall activity in the network.  Also, agents who benefit most from the activity choose to have more interactions, further influencing their behavior and that of others.  As I show below, these two effects lead to systematic and predictable overall distortions in the equilibria of such games played on networks.  This predictable pattern allows me to derive welfare implications of the friendship paradox.

In order to study peer effects in the presence of the friendship paradox, I start by introducing a Bayesian version of a linear-quadratic game of strategic complements.  This extension of the popular network game is useful because in it agents' behaviors can solved for in closed-form as a function of their types and degree.  This variation of a network game should be useful beyond the current paper.

In this game,  agents' behaviors are ordered by their degree.   This leads to greater equilibrium behavior in a game on a network compared to a benchmark society with uniform-at-random matching, due to the fact the added influence of
friends in a network who have higher degrees (in the sense of first-order stochastic dominance) than uniform at random matchings from a population.  Next, I endogenize the network, showing that people with greater preference for an activity choose to have higher degree, and that this leads to a further amplification of the overall activity in the society.
I also provide results on comparative statics and welfare orderings, showing how the friendship paradox improves overall welfare in settings with positive externalities and is harmful in settings with negative-enough externalities.

I then show that the results extend to general settings with complementarities by adapting results on monotone comparative statics to the current setting.

Finally, in the appendix, I examine how the results change when moving to a setting of strategic substitutes.  There, higher degree agents (on a fixed network) engage less in a behavior when they are exposed to more of the behavior by their friends.  This leads agents in a network to perceive less behavior by their neighbors, compared to a random matching, as the highest degree individuals exhibit the least activity.  Given the substitute condition, agents respond to less perceived behavior by their neighbors by engaging more themselves.  Thus, when accounting for the behavior as a function of degree and its feedback, activity on a network ends up being greater in the network setting than in a benchmark with uniform-at-random matching.
However, once the network is endogenized, the result in the case of strategic substitutes becomes ambiguous, as then agents with greater preference for the activity choose higher degree, but this offsets the proclivity of higher degree agents to free-ride more, leading to competing effects.
Thus, while the ordering in the context of endogenous networks with strategic complements is clear and amplifies behavior of all agents, in the case of strategic substitutes the overall effect is ambiguous.

\section{A Model and the Friendship Paradox}

\subsection{Agents and Random Networks}

A finite set $N=\{1,\ldots, n\}$ of agents, with generic indices $i,j$, are members of a
random network.

We examine interactions at an interim stage, when each agent knows his or her degree but not the full structure of the network.%
\footnote{For more on this perspective, see \citet{jacksony2005,jacksony2007,manshadij2009,galeottigjvy2010}.}
In particular, agents do not know how many friends each of their friends has (or will have).

Agent $i\in N$ has $d_i\in \{1,2,\ldots\}$  links in the network.\footnote{Agents who are isolated play no role in what follows, and so I focus on the population of agents who have at least one connection in the network, and so $d_i$ is always positive. }

Let $P_{i}(d)$ denote the degree distribution of the population of $i\in N$'s potential neighbors under a random network formation model,
not conditioning on the fact that the person ends up connected to $i$;
and suppose that this marginal distribution is the same
for each of $i$'s neighbors.
Let $\E_{i}[\cdot]$  denote the expectation and $\Var_{i}[\cdot]$ be the variance associated with $P_{i}$.

I am deliberately vague about the specifics of the network formation model, since this approach embodies essentially all of standard random network models: what is important
is that we can quantify agents' beliefs about their neighbors' degrees.
For example, the joint distribution over neighbors' degrees in most network formation models involve correlations across neighbors, which is fine here.
All we need to track for our analysis are the marginal distributions.
Thus, this allows for general degree distributions, including scale-free distributions, Poisson distributions, and hybrids.
The only condition on the network formation process presumed in what follows is that any idiosyncracies (e.g., homophily, assortativity, etc.) in the distribution are already accounted for in subscripting by $i$, which
can thus condition on $i$'s characteristics and degree, and
then the relative chance that one of $i$'s links goes to a neighbor with degree $d$ will be in relative proportion to that degree.

The probability that some given link of $i$ connects to an agent who has degree $d$ is given by
\begin{equation}
\label{tilde}
\widetilde{P}_{i}(d) = \frac{d}{\E_{i}[d]} P_{i}(d) .
\end{equation}

Let me emphasize the perspective here.  A network has formed, or will form, and we examine a particular node $i$ who knows its degree $d_i$ and
the distribution from which the degrees of its neighbors are drawn but not their actual degrees.   The degrees of the potential neighbors are described by $P_i$.   If we look at any one of $i$'s links and ask what the  distribution of degrees of the neighbor on that link is, then it is described by $\widetilde{P}_{i}(d)$.\footnote{For more discussion of this, see \citet*{newmansw2001} and Section 4.2 of \citet{jackson2008}.}
This follows directly since people with higher degrees must be friends more frequently -- in proportion to their degree.  For instance, if half of the population has degree 2 and half has degree 1, then two thirds of the friends in the network must be of degree 2 as they are twice as likely to be linked to as the degree 1 people.

\subsection{The Friendship Paradox}

Let $\widetilde{\E}_i$ denote expectations with respect to $\widetilde{P}_{i}$.
From (\ref{tilde}) it follows that the expected degree of $i$'s neighbors (the expectation of $d$ under $\widetilde{P}_{i}(d) $)
is
\begin{equation}
\label{etilde}
\widetilde{\E}_i[d] = \sum_d  d \widetilde{P}_{i}(d) = \sum_d  d \frac{d}{\E_{i}[d]} P_{i}(d) =  \frac{\E_{i}[d^2]}{\E_{i}[d]} =  \E_{i}[d] + \frac{\Var_{i}[d]}{\E_{i}[d]}.
\end{equation}

This leads to the following lemma, which is a more explicit statement of the friendship paradox in the context of such a random network model.

\begin{lemma}
\label{friendshipparadox} [The Friendship Paradox]

The expected degree of a neighbor of any agent $i$ is $\widetilde{\E}_i[d] = \E_{i} [d] + \frac{\Var_{i}[d]}{\E_{i}[d]}$.
\end{lemma}

In an extreme case, in which all nodes are perfectly positively assortatively matched, then $\Var_{i}[d] =0$ and the expected degree of a node's neighbor is simply the same as its degree.
However, generally there is some variation in degree across neighbors and so the paradox implies that the average degree of the population of potential neighbors will be strictly less than the average
realized neighbors' degree.

In particular, if the expectations, $\E_i$, are similar across agents and we can drop the subscript, and the variance is positive (the network has some possibility of not being regular), then we get an immediate corollary that
\[
\E[d]<\widetilde{\E}[d] = E [d] + \frac{\Var[d]}{\E[d]}.
\]
So, the expected degree of a neighbor is the population average plus a factor which is the ratio of the variance of the distribution over the average.
Moreover, all agents whose degrees are no higher than average, or in fact are even slightly above average, have strictly lower degrees than the expected degrees of
their neighbors.  It is only agents whose degrees are substantially above the average (by at least $\frac{\Var[d]}{\E[d]}$) who have degrees as high as
their neighbors' expected degrees.

\section{A Linear-Quadratic Game}

Let us now analyze the impact of the friendship paradox in the context of a setting with strategic complementarities.
We first explore how the friendship paradox plays out in a linear-quadratic setting since it admits a closed-form solution and cleanly illustrates the intuition
behind the general results.
This is a variation on the games studied by
\citet*{ballestercz2006,bergemannm2013,bramoullek2014,belhajbd2014,demartiz2015}.\footnote{For an overview, see \citet*{jacksonz2014}.}

I study a variation in which agents choose actions based on expected values of friends' actions.
This differs from the analysis in \citet*{ballestercz2006} which analyzes a complete information game.
The incomplete information version
allows us to quantify actions by degrees.  In a complete information game, each agent's actions are dependent upon the full network structure, which then leads to challenging comparative statics (e.g., see the discussion in \citet{jacksony2005,jacksony2007,galeottigjvy2010}).

Thus, the equilibrium characterizations here may be of independent
interest beyond the analysis presented in this paper, since they provide a simpler closed form for actions as a function of type and degree.%
\footnote{As degree is strongly correlated with other centralities, such as the Katz-Bonacich centrality which characterizes behavior in \citet{ballestercz2006}, this can also be thought of as a first-order approximation.
Beyond this, previous studies have found that people know remarkably little beyond their immediate
neighborhoods, and are not well-informed about friends of friends who are not already direct friends (e.g.,
see \citet*{friedkin1983,krackhardt1987,krackhardt2014}).
Furthermore, networks change over time and people are often choosing behaviors
 in reactions to their perceptions and anticipations of their future interactions.  Those anticipations may be based on past experiences, which tend to be biased towards higher degree individuals via the friendship paradox.   Thus, a Bayesian equilibrium in a random network may in fact be a better approximation than a complete information Nash equilibrium in some snapshot of a network for many applications.}

\subsection{Three Games}

The analysis lies in the comparison of three closely related games.   Each game is based on the same setting, but differs in terms of whether agents make decisions based on their friends' behaviors or the overall population's behavior, and whether the utility that they ultimately experience depends on their friends' behaviors or the overall population's behavior.

Thus, following the terminology of \cite{khanemant2006}
 it is important to distinguish between the `decision utility' that agents perceive when making choices, and the `experience utility' that they ultimately receive once the game is played.   As Khaneman and Thaler point out, decisions could be made based on information from past experiences, which could differ
  from what leads to current payoffs.  In our application,  past interactions that lead to the decision utility would be based on interactions with friends,
 while the future utility that agents experience could be based on societal norms.  This shows the impact of the friendship paradox.

 Thus, to understand the effect, I define three games:  (i) a benchmark in which both decision and experience utility are based on society-wide behavior, (ii)
 another game that agents may perceive that they play in which both decision and experience utility are based on interactions only with friends, and (iii) the game that captures the paradox in which decisions are based on friends' behavior but experience utility depends on society-wide behavior.

In all of the games, agent $i$ has a `type' $\theta_i\in \Theta$, where $\Theta$ is a compact subset of $\Re_{+}$.  Types have a support that includes positive values,
and may be correlated with degrees (as we will derive in the endogenous network case below).  So, extend $P_i$ and let $\widetilde{P}_i$ denote the probability  distributions
that $i$ perceives jointly over the types and degrees
of her potential neighbors (unconditionally and conditional upon being linked, respectively); and so when using $P_i$ and $\widetilde{P}_i$ in the previous section we were considering its marginal just on degrees.

\subsubsection{The `Society-Wide' Game}

The first game is the basic one in which each agent cares about how his or her behavior matches with a society-wide norm ${\E}_i\left[  x_j \right]$, where $j$ is a generic
other member of the population.

Agent $i$ chooses an action $x_i\in \Re_+$ and gets expected utility  described by
\begin{equation}
\label{eusoc}
EU_i^{soc}(x_i,d_i,\theta_i,x_{-i})\equiv \theta_i x_i + a  x_i  d_i {\E}_i\left[  x_j \right] - \frac{c x_i^2}{2} +  \phi \E_i\left[ \sum_{j\neq i} x_j\right].
\end{equation}
The agent's type $\theta_i$ determines his or her idiosyncratic taste for the action.
The scalar $a>0$ captures the level of complementarity of the agent's action with the actions of others,   $c>0$ scales the cost of taking the action, and $\phi \in \Re$ is a parameter that captures the extent of global externalities - either positive or negative.
For instance, if $x_i$ is a level of criminal activity then $\phi$ would be negative representing societal costs of crime, while if $x_i$ is a level of knowledge acquisition or human capital investment
then $\phi$ would be positive. 

There is a unique Bayesian equilibrium to this game when agents simultaneously decide upon an action $x_i$ as a function
of their type and degree, solved for explicitly in Lemma \ref{eq} below.

Let $x^{soc}(\theta_i,d_i)$ denote the equilibrium action, and let
\[
U^{soc}(\theta_i,d_i)\equiv EU^{soc} (x^{soc}(\theta_i,d_i),\theta_i,d_i; x^{soc}(\theta_{-i},d_{-i})),
\]
denote the equilibrium expected utility in the society-wide equilibrium.

\subsubsection{The `Friend' Game}

The second game is a variation of the game in which agents' payoffs depend on a complementarity between their action and those of their friends.
Thus, they maximize
\begin{equation}
\label{eu2}
EU_i^{friend}(x_i,d_i,\theta_i,x_{-i})\equiv \theta_i x_i + a  x_i  d_i \widetilde{\E}_i\left[  x_j \right] - \frac{c x_i^2}{2} +  \phi \E_i\left[ \sum_{j\neq i} x_j\right].
\end{equation}
Note that the first expectation is over $i$'s friends (indexed by $j$), so $i$ conditions on being linked to them and hence the $\widetilde{\E}_i$, while the second expectation is over the whole population as it is a global externality
and so is simply $\E_i$.

Again, as shown in Lemma \ref{eq} below, there is also a unique Bayesian equilibrium to this game.
Let $x^{friend}(\theta_i,d_i)$ denote the equilibrium action choice and let
\[
U^{friend}(\theta_i,d_i)\equiv EU^{friend} (x^{friend}(\theta_i,d_i),\theta_i,d_i; x^{friend}(\theta_{-i},d_{-i}))
\]
denote the equilibrium expected utility in the friend game.

\subsubsection{The `Naive' Game}

In the naive game, agents misperceive the game.   They form expectations based on their perceptions of their friends' actions and so make decisions to maximize
$EU_i^{friend}$, but actually experience utility according to $EU_i^{soc}$.   The interpretation is that they don't understand the friendship paradox and so take their
friends' behaviors to forecast the society-wide behavior.

In this case, the equilibrium will again be described by the behavior $x^{friend}$, since agents think that they have expected utility described by $EU_i^{friend}$.

We are interested in the utility that these ``naive'' agents actually experience.
The expected utility that they experience is described by
\[
U^{naive}(\theta_i,d_i)\equiv EU^{soc} (x^{friend}(\theta_i,d_i),\theta_i,d_i; x^{friend}(\theta_{-i},d_{-i})).
\]

The comparison of this utility to that from the society-wide game is relevant to understanding how the friendship paradox increases norms of behavior and distorts
welfare, and is important in understanding certain policy designs (see Section \ref{policy}).

\medskip

Let me make one comment on the payoff structure before proceeding.
In (\ref{eusoc}) and (\ref{eu2}) people get utility according to the anticipated {\sl average} actions of others times their own degree -- so average behaviors are what they care about, but people with
higher degrees have stronger complementarities.
It is not necessary for the results that degree enter linearly - the full generality of the results is made clear in Section \ref{generalcase}.  Nonetheless, it is essential that people with higher degrees have higher complementarities.\footnote{See the references in \citet{jacksonz2014} for a variety of payoff structures for games on networks.}  For instance, having more opportunities to drink or consume drugs, where degree represents how many parties a person attends, can lead to more incentives to undertake the behavior.    If degree is completely dropped from the payoff function then the friendship paradox will not play a role as people's behavior would be uncorrelated with their degrees.

\subsubsection{Equilibrium}

To solve for equilibrium in closed form,  in the analysis of the linear-quadratic game I restrict attention to the case in which all agents face the same degree distribution over their neighbors' types and degrees; so $P_i$ is the same for all $i$.
Let $\widetilde{\E} \left[ \cdot \right]$ denote $ \widetilde{\E}_i\left[  \cdot_j \right]$, for a generic $i$.
Agents may differ with regards to their own realized type and degree, but their expectations over the rest of the population are
similar.

I consider the Bayesian equilibria in which agents choose actions simultaneously.

Throughout the analysis of the linear-quadratic setting, I also maintain the assumption that $c> a \widetilde{\E}\left[  d \right] $.  This ensures that an equilibrium to each of the games exists, as otherwise the
incentives from complementarities dominate the negative incentives from the costs and iterative best responses diverge.

\begin{lemma}
\label{eq}[Equilibrium Characterization]\  \
There is a unique Bayesian equilibrium in each game:\footnote{Comparative statics are provided in the appendix.}
\begin{equation}
\label{eqxsoc}
x^{soc}(\theta_i,d_i) \ = \ \frac{\theta_i }{c}+ \frac{a d_i  {\E}\left[\theta\right] }{c\left(c -  a  {\E}\left[  d \right]\right)  },
\end{equation}
\begin{equation}
\label{eqx}
x^{friend}(\theta_i,d_i) \ = \ \frac{\theta_i }{c}+ \frac{a d_i  \widetilde{\E}\left[\theta\right] }{c\left(c -  a  \widetilde{\E}\left[  d \right]\right)  }.
\end{equation}
\end{lemma}

\subsection{Behavior and Welfare with Exogenous Interactions}

To compare behavior and welfare across games,  let us
first consider cases in which $\theta_j$ and $d_j$ are uncorrelated so that the expectation of a neighbor's $\theta_j$ is simply
a random draw from the populations' distribution of $\theta$'s:
$\widetilde{\E}[\theta_j] = \E[\theta_j]$.\footnote{Thus, the joint distribution $\widetilde{P}$ on a neighbor's $\theta_j,d_j$ is a product of $P$ on $\theta_j$
and $\widetilde{P}$ on $d_j$.}
This allows us to separate the effects of the degree of the agent and their preference types.

I refer to this as the case with `exogenous' interactions.  This contrasts
with `endogenous' interactions (Section \ref{endogamp}) in which agents choose their number of interactions $d_j$ after knowing their
 preference type $\theta_j$.   In that case, the $\theta_j$'s determine the $d_j$'s and the two end up correlated.

\subsubsection{Comparing Behavior Across Games}

The first main result is that if agents react to their friends' behaviors -- regardless of whether that is because they only care about their friends' behaviors or
because they mistakenly use their friends' behaviors to forecast the society-wide norm -- then their behavior is greater than if
they best respond to society-wide behavior.

\begin{proposition}
\label{comparison}
[The Impact of the Friendship Paradox on Behavior] \  \
Consider a random network with a degree distribution that has positive variance and
for which  $\widetilde{\E}[\theta_j] = \E[\theta_j]$.
Then, $x^{friend}(\theta_i,d_i)> x^{soc}(\theta_i,d_i)$ for all $\theta_i$ and $d_i$.  Thus,
$\widetilde{\E}[x^{friend}] >\E[x^{friend}]  > \E[x^{soc}]$.
\end{proposition}

Proposition \ref{comparison} states that equilibrium behaviors of all types of agents are strictly higher when they are interacting in a network and exposed to the
friendship paradox, as compared to being randomly matched to the population without weighting by degree.  It also states that expected neighbors' behaviors are even higher than the population average under the network equilibrium.  This last observation is really what drives the result: neighbors in a network have higher expected degree than the population average and so are expected to be more engaged in the behavior given the complementarities.  Greater activity by neighbors feeds back via the complementarities and raises the overall equilibrium behaviors in the network compared to the population-matching benchmark.

\begin{figure}[h!]
\centering
{
\label{fig:ratios}
\includegraphics[width=0.6\textwidth]{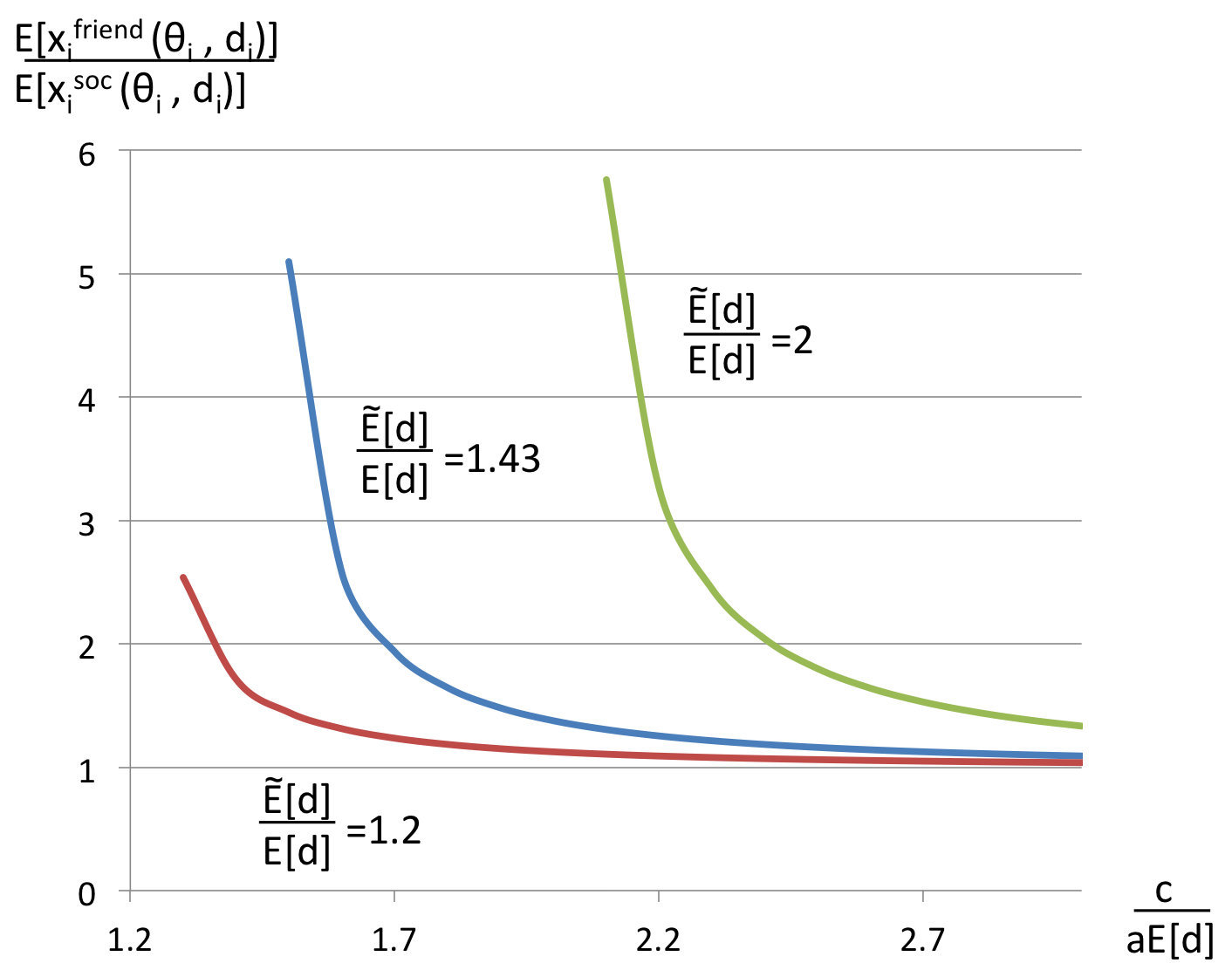}
}
\caption{\label{ratios}  The ratio of $x^{friend}/x^{soc}$ evaluated at the average degree and type.   The $x$-axis is the value of $c/(a \E[d])$ and the three different curves correspond to three different values of $\widetilde{\E}[d]/\E[d]$, with the 1.43 version from the Indian village data, and the other two for higher and lower ratios.}
\end{figure}

Figure \ref{ratios} shows the magnitude of the network effect as a function with parameters, plotting $x^{friend}/x^{soc}$ as a function of the background parameters.
 The parameters can be collapsed to one parameter $c/(a\E[d])$, capturing the
relative cost of action compared to the social interaction factor and average degree.  Three curves are plotted for different values of how different the average
degrees of neighbors are compared to the average degree.
Utility comparisons are even more dramatic as actions enter quadratically (own times others' actions) - so those would correspond to a shift of these curves squared.

\subsubsection{Comparing Welfare Across Games}

The ranking of equilibrium actions has strong welfare implications: we can Pareto rank the utilities from the various games, depending
on the nature of the global externalities.

Our second main result compares the experienced welfare between the case in which agents mistakenly use their friends to forecast social norms,
$U^{naive}$, and the
case in which they correctly forecast the social norm,  $U^{soc}$.    (The comparison with  $U^{friend}$ is included in the appendix.\footnote{%
It is clear that $U^{naive}$ is always worse than $U^{friend}$ since the global externalities are the same but agents lose some of the local interaction effect due to
playing with society rather than the expected inflated behavior under the friendship paradox.  People are also always disappointed since they are not best-responding to the distribution of actions that eventually determine the payoff that they experience.
The more nuanced comparison is between $U^{naive}$ and $U^{soc}$, as captured in Proposition \ref{welfare2b}.})

Whether the inflated behavior due to the friendship paradox improves or harms welfare depends on the nature of the global externalities:  it improves welfare when those externalities are nonnegative but harms welfare when externalities are negative enough.

\begin{proposition}
\label{welfare2b}
[Strict Pareto Rankings and the Friendship Paradox] \  \
Consider a random network model with a degree distribution that has a positive variance and
for which $\widetilde{\E}[\theta_j] = \E[\theta]$.
If $E(\theta)< 2a E[d]/c$ and global externalities are positive or not too negative (there exists $ \overline{\phi}<0$ such that if $\phi \geq \overline{\phi}$),
then:\footnote{%
$U^{naive}(\theta_i,d_i) > U^{soc}(\theta_i,d_i) $ holds without the condition $E(\theta)< 2a E[d]/c$ if global externalities are positive enough.  }
$$ U^{naive}(\theta_i,d_i) >
U^{soc}(\theta_i,d_i)$$
for all $\theta_i$ and $d_i$.
If global externalities are negative  enough (there exists $\underline{\phi}< \overline{\phi}$ such that if $\phi \leq \underline{\phi}$),
then:
$$U^{soc}(\theta_i,d_i)   > U^{naive}(\theta_i,d_i)$$
for all $\theta_i$ and $d_i$.
\end{proposition}

The friendship paradox amplifies behaviors, and thus if global externalities are positive (or not too negative) this actually improves overall welfare.  People do not take their externalities into account, and so the inflated behavior due to the friendship paradox is actually helpful if global externalities are not too negative.
In contrast, if externalities are negative enough then the inflated behavior is harmful.

The reason for a gap between $\underline{\phi}$ and $ \overline{\phi}$
is that there are two forms of
externalities.
Local externalities are always positive and come through the complementarities of the actions of the agents, and global ones which could be positive or negative.  In the case where both forms of externalities are positive, then more behavior by other agents strictly increases an agent's payoff from any given level of behavior, and thus from the best response too.  In that case, there is strictly more activity by neighbors in the network setting than in the benchmark and so each agent of any type gets a higher utility from any level of her behavior, and thus also when comparing best responses.
Once the global externality is negative enough, agents' utilities are hurt so much by others' behavior, that even the benefits that they see from the local externality cannot offset the loss, and in that case they prefer to have less engagement in the behavior by all agents and so prefer all correctly responding to the society-wide norm than having
behaviors amplified by the friendship paradox.
In the region between $\underline{\phi}$ and $ \overline{\phi}$, people with higher types and degrees benefit enough from the feedback due to the friendship paradox to overcome the negative global externalities and prefer the network setting, while people with lower types and degrees do not and prefer the benchmark setting.

Generally, all three games have Pareto inefficient actions, except for knife-edge cases, since agents are only maximizing their own utilities and not taking into account the externalities that their actions have on others.
Nonetheless, this shows that the friendship paradox has strong welfare implications compared to what would happen without networked interactions.
In cases such as investing in education or human capital (e.g., studying), which have positive externalities, the fact that people may base their choices off of popular individuals who have more incentives to invest in human capital is welfare-enhancing.
In contrast, in cases such as delinquent behaviors among teens which have substantial negative externalities, the friendship paradox decreases welfare.

\subsubsection{Increased Inequality in Utility}

Beyond the main results for the exogenous case that are given above, we can also show that
the friendship paradox increases the inequality in actions and welfare among the population.  This is captured in the following
proposition (see also Proposition \ref{inequality2} in the appendix for additional results).

\begin{proposition}
\label{inequality}
[Increased Inequality] \ \
Consider a random network model $(a,c,\phi, P)$, that has a degree distribution that has a positive variance and
for which $\widetilde{\E}[\theta_j] = \E[\theta]$.
Consider any $i$ and two different degrees $d_i>d_i'$.
Then
\[
x^{friend}(\theta_i,d_i) - x^{friend}(\theta_i,d_i')  > x^{soc}(\theta_i,d_i) - x^{soc}(\theta_i,d_i').
\]
Also, if $E(\theta)< 2a E[d]/c$ then
\[
U^{naive}(\theta_i,d_i) - U^{naive}(\theta_i,d_i') > U^{soc}(\theta_i,d_i) - U^{soc}(\theta_i,d_i')
\]
for all $\theta_i$.
\end{proposition}

The proposition states that the friendship paradox increases the inequality in actions and payoffs between more and less popular/central (higher versus lower degree) individuals.
The intuition behind this proposition is that agents benefit from the interaction with their neighbors, and higher degree people enjoy greater interactive effects.
Since the friendship paradox produces larger expected actions of neighbors, this increases the difference in the local externality experienced by higher versus lower degree individuals, which affects both actions and payoffs in the same direction.
The difference in global externalities is the same regardless of degree, so those are inconsequential.\footnote{%
The same comparison follows if instead of comparing the friendship actions or utility to the benchmark, one compares settings with more social interactions (e.g., a first-order stochastic dominance shift in $\widetilde{P}$ or an increase in $a$ or lower $c$) to one with less social interactions.  This is shown as Proposition \ref{inequality2} in the appendix.}

This means that as a society experiences technological changes that enable greater social interaction, then for behaviors that involve complementarities there will be an increase in inequality in behavior and welfare between agents who have more interactions and those who have fewer.
This provides a very different lens into inequality than other discussions in the literature: it is not a point about income or wealth, but just about utility - which has clearly ordered comparative statics in this model.

 \subsection{Endogenous Interactions and Further Amplifications of Behavior}\label{endogamp}

We have seen that complementarities lead agents with higher degrees to engage more in a behavior,
 which then feeds back to lead to further amplify the behavior of all agents given that higher degree individuals have a disproportionate
influence on others' behaviors.

The next main result outlines how endogenizing the network amplifies the effect of the friendship paradox.
In the above analysis we treated preference types and degree as independent.   Once
the network is endogenized, preference types and degrees necessarily become positively correlated.
People with more taste for the behavior -- agents with higher $\theta_i$'s -- benefit
more from the interactions with others, and thus prefer to have a higher degree.
Thus, when we model network formation  we see a positive correlation between $\theta_i$ and $d_i$.
This leads agents with higher degree to engage even more in the activity, further increasing the effect of the friendship paradox.

To understand this effect, note that people who get more enjoyment from some interactive behavior (especially in a social context, for instance adolescents with a predisposition for drugs or alcohol) will prefer to interact more.
This selection effect further increases the behavior of agents with high degrees, as they benefit not only from the increased complementarity that accompanies their high degrees, but they also tend to have higher
base propensities for the behavior to begin with.  This  amplifies  the feedback and thus behaviors throughout the population.

Consider a game in which agents choose $d_i\in \{0,1,\ldots,n-1\}$ in a first stage (as a function of their $\theta_i$'s), and 
anticipate playing the friendship game and choosing $x_i$ in a second stage.
Also, consider the case in which the choices of $d_i$'s are private, so that agents do not observe others' choices (given that they do not see their neighbors' degrees).
The game in which those choices are public has similar results but seems less natural, and this formulation allows the use of Bayesian equilibrium rather than perfect Bayesian; but there do not appear to be any interesting strategic differences between the games.

Forming relationships is costly, and in order to obtain a closed-form solution, consider a quadratic cost function of the form $C(d_i)= \frac{k d_i^2}{2}$ for $k>0$.

Let the distribution over types be i.i.d. across agents and be atomless with compact support.  This ensures that any symmetric equilibrium is essentially in pure strategies,
as at most a set of measure 0 of types ever have multi-valued best responses.\footnote{Given that the set of agents who ever are indifferent are of measure 0, any equilibrium that involves mixing has a counter-part
in which the indifferent agents do not mix (and their decisions do not affect any other agents' best responses given their negligible measure) and the equilibrium still results in the same distributions over degrees and actions.}  The extension to distributions with atoms is straightforward since degree choices are still nondecreasing (in the sense of first-order stochastic dominance)
and each type mixes over at most two adjacent degrees.

Let $d^{end}(\theta_i)$ be the (Bayesian) equilibrium degree choice of an agent of type $\theta_i$ and
$x^{friend}_{end}(\theta_i,d^{end}(\theta_i))$ be the resulting equilibrium actions in the second period, in the overall endogenous-network equilibrium.%
\footnote{The notation indicating the dependence of $x^{friend}_{end}(\theta_i,d^{end}(\theta_i))$ on $d^{end}(\theta_i)$ is redundant as they are both tied down in equilibrium as a function of $\theta_i$, but this notation will also be useful in comparing actions
to the benchmark.}

For tractability, I ignore that agents might sometimes have their degrees adjusted slightly from their desired levels.   For instance, if $n=3$ and $d_1=1$ while $d_2=d_3=2$, then there is no network that can give all of the agents their preferred degrees at the same time.  So, some agent would have to be rationed by a link.   Including the probability of being rationed would render the calculations intractable without adding any insight.
For large $n$ the probability that any agent would have to be rationed would go to 0, presuming some simple bounds on preferred degrees, in standard random network models such as the configuration model.  Thus, I ignore this effect and allow agents to unilaterally decide on their degree, but the model could be extended to use the configuration model and account for agents' expectations that their degree is not exactly realized.  This would prevent solving for the equilibrium in closed form, but would still lead to qualitatively similar results.

Thus, agents' choices of $d_i$ and $x_i$ maximize their expected utility, anticipating equilibrium choices of the other agents.
In particular, equilibrium choices are solutions to the problem
\begin{equation}
\label{eud}
\max_{d_i,x_i} \ \theta_i x_i + a   x_i d_i \widetilde{\E}\left[  x^{friend}_{end}(\theta_j,d_j(\theta_j)) \right] - \frac{c x_i^2}{2} -  \phi (n-1)\E\left[ x^{friend}_{end}(\theta_j,d_j(\theta_j))\right] - \frac{k d_i^2}{2},
\end{equation}
where expectations $\widetilde{\E}$ are relative to the equilibrium distribution of others' choices.

Symmetric pure-strategy equilibria exist,\footnote{\label{costs}One needs that $c>a \widetilde{\E}[d]$ for the second stage to be well defined. This last condition refers to an endogenous parameter, which can be checked ex post. A sufficient condition is simply that $c>a(n-1)$, which I  maintain to ensure existence of equilibrium for all possible degree choices (which are bounded by $n-1$).  Similarly, I assume that $k>a(n-1)$, which ensures concavity of payoffs in degree (as shown in the appendix). }
and as there can be multiple equilibria I provide results that hold for any symmetric pure-strategy equilibrium with at least two degrees.\footnote{The game is not quite supermodular, as actions depend both on types and degree.   For example, if one increases the degree that a low type chooses, then that increases the probability that a friend is of a low type, which can decrease expected neighbors' actions.
Nonetheless, best responses are still monotone, and essentially unique and can be taken as step functions with at most $n$ values, and so can be taken to be compact in the weak topology, and then existence is easy to establish from standard arguments.}$^,$\footnote{Here the equilibrium may not be unique. For instance, there always exists an equilibrium in which all agents choose $d_i=0$ given that they expect all others do as well.  But the characterization here
applies to equilibria in which some agents choose a positive degree.  Such equilibria exist by standard arguments when restricting degree choices to be positive, and then such equilibria remain an equilibrium without that restriction for $k$ that are not too overwhelming.  }

To compare actions, it is useful to compare the situation in which the degree correlates with the type to the situation in which we maintain the same distribution over degrees and over types, but in which we take them to be independent.
In particular, let $x^{friend}_{\perp}$ denote the actions in the friend game when the marginal distributions over degrees and types are the same as under the endogenous equilibrium, but they are taken to be independent so that the joint distribution is the product distribution.
Similarly, define $x^{soc}_{end}$ and $x^{soc}_{\perp}$ be the equilibrium actions in the society-wide game taking $d^{end}$ as defined above (from the friend game) as given.
Note that from Lemma \ref{eq}, $x^{soc}_{end}=x^{soc}_{\perp}$.

\begin{proposition}
\label{endogenous}[Endogenous Network Amplifications]
\  \
Consider an endogenous symmetric pure-strategy equilibrium of the endogenous friend game, $d^{end}$,
and suppose that it involves at least two degrees.
Then each agent chooses a degree which is a nondecreasing function of the agent's type.
Moreover, for all $\theta_i$ for which $d^{end}(\theta_i)>0$:
$$x^{friend}_{end}(\theta_i,d^{end}(\theta_i))> x^{friend}_\perp(\theta_i,d^{end}(\theta_i))  > x^{soc}_{end}(\theta_i,d^{end}(\theta_i))= x^{soc}_{\perp}(\theta_i,d^{end}(\theta_i)).$$
\end{proposition}

Proposition \ref{endogenous} distinguishes the two effects that we have been discussing.  First, having higher degree people be neighbors more often leads them to have more influence, and their tendency for more engagement in the behavior given their higher rate of interaction leads to greater behaviors by agents of all types in response.
This is the general comparison between $x^{friend}_\perp$ and $x^{soc}$, which follows from before.
Second, higher-type agents benefit more from having higher degrees leading to a positive correlation between degree and type, further increasing high-degree agents' behaviors and further increasing the behaviors of all agents.  This is the comparison between $x^{friend}_{end}$ and $x^{friend}_\perp$.
Given these rankings of activity for each type and degree, it follows directly that the average rankings follow the same rule.

In these comparisons I have worked with $d^{end}$ from the endogenous equilibrium in the friend game.   One could also instead use a different endogenous
equilibrium for each version of the game.  The complementarities are greatest in the
friend game, and with the correlated type-degree distribution.  Thus, in a sense, the proposition above is even more demanding, as it holds
even when the other equilibria are evaluated at the greatest degree distribution.\footnote{The possibility of multiple equilibria means that one has to select comparable equilibria across games - for instance the maximum equilibrium.   Also, even if one gets a degree distribution that
 first-order stochastically dominates another, it does not necessarily mean that the same is true of $\widetilde{P}$.  Nonetheless, the comparison with $x^{soc}$ does
 not require that as it would be between the friend distribution for the endogenous friend game and the society distribution for the endogenous society game, and so the comparison between $x^{friend}_{end}$ and $x^{soc}_{end}$ would be preserved.}

There are implications for welfare that are analogs of Proposition \ref{welfare2b} (and Proposition \ref{welfare2}):   the ordering of the
behavior correspond to the size of the externalities, and so with positive externalities the further amplification of the endogenous interactions increases welfare, while for negative enough externalities it decreases welfare.

\section{General Games with Complementarities}\label{generalcase}

I now show that the results above extend to general games of strategic complements without the linear-quadratic structure.\footnote{%
Omitted definitions here are standard from the literature on supermodular games (e.g., see \citet{milgroms1994,vanzandtv2007}), but I adapt the structures to
a network setting.  For more background on some related games, see \citet*{jacksony2005,jacksony2007,sundararajan2007,jackson2008,manshadij2009,galeottigjvy2010}.}
A discussion of strategic substitutes appear in the appendix.

Each agent $i$ chooses a strategy from a set $X$, which is a compact metric lattice with associated partial order $\geq_X$.
For each $i$, let $\theta_i$ lie in a partially ordered set $\Theta$ with associated partial order $\geq_\Theta$.
The utility of agent $i$ with degree $d_i$ of type $\theta_i$ when other agents play actions $x_{-i}$ is given
by
\[
u_i( x_i; (x_{j})_{j\in N_i}, \theta_i, d_i),
\]
where $N_i$ is the realized set of neighbors of $i$.\footnote{We could also allow for global externalities as a function of $x_{-i}$.  For simplicity, I drop that notation, but the results below apply directly, just with the additional notation, since $N_i$ is defined to be the set of other agents whose actions interact strategically with $i$'s action.
}
The Bayesian version of this game will just expect over the neighbors.

An agent's payoff to choosing some action $x_i$ depends on
the action of the agent's neighbors, the agent's type $\theta_i$, and the agent's degree $\left((x_{j})_{j\in N_i}, \theta_i, d_i\right)$.

In order to define complementarities in the general setting, we need to be able to discuss how an agent's payoffs from taking various actions
change when his or her circumstances $\left((x_{j})_{j\in N_i}, \theta_i, d_i\right)$ ``increase''.

What it means for $\left((x_{j})_{j\in N_i}, \theta_i, d_i\right)$ to ``increase'' is mostly obvious except for the fact that as we increase an agent's degree then the set of neighbors that the agent has changes.
The requirement that I work with is simply that the set of old neighbors' actions do not decrease.

Let us say that $\left((x_{j}')_{j\in N_i'}, \theta_i', d_i' \right) \geq \left( (x_{j})_{j\in N_i}, \theta_i, d_i\right)$
if $\theta_i' \geq_\Theta \theta_i$,
$d_i'\geq d_i$,
$N_i\subset N_i'$, and
$x_{j}'\geq_X x_{j}$ for each $j\in N_i$.
So, when we add neighbors to an agent's set of friends, the agent's circumstances are `weakly greater' if the old neighbors' actions have not decreased.

With this partial ordering defined for the network setting,
we can then define $u_i$ to have {\sl increasing differences} in $( x_i; (x_{j})_{j\in N_i}, \theta_i, d_i)$ in the usual manner:
\[
u_i( x_i'; (x_{j})_{j\in N_i}, \theta_i, d_i)- u_i( x_i; (x_{j})_{j\in N_i}, \theta_i, d_i)
\]
is nondecreasing in $(x_{j})_{j\in N_i}, \theta_i, d_i$ when $x_i'\geq x_i$.

Increasing differences is the standard definition adapted to a network setting.  It asks that if we increase an agent's type, number of friends, and the behaviors of the agent's neighbors, then the agent's relative preference for  greater behavior (weakly) increases.

Some examples of utility functions that satisfy increasing differences are
\begin{itemize}
\item [1.] $u_i( x_i; (x_{j})_{j\in N_i}, \theta_i, d_i) = \theta_i x_i + a  x_i f\left(\sum_{j\in N_i} x_j\right) - \frac{c x_i^2}{2} $ for any increasing $f$ (including strictly concave $f$'s),
\item [2.] $u_i( x_i; (x_{j})_{j\in N_i}, \theta_i, d_i) = v(x_i;\theta_i,\max_{j\in N_i}[x_j])$  and $v(x_i;\theta_i, m)$ satisfies increasing differences in $(x_i; \theta_i,m)$,
\item [3.] $u_i( x_i; (x_{j})_{j\in N_i}, \theta_i, d_i) = v(x_i; \theta_i,T((x_j)_{j\in N_i}))$, where $T(\cdot)$ is the number of elements in $(x_j)_{j\in N_i}$ exceeding
  some threshold $t\in X$ and $v(x_i;\theta_i,T)$ satisfies increasing differences in $(x_i;\theta_i,T)$,
\item [4.] $u_i( x_i; (x_{j})_{j\in N_i}, \theta_i, d_i) = v(x_i; \theta_i, g((x_j)_{j\in N_i}))$, where $g:(X\cup X^2\ldots \cup X^{n-1})\rightarrow \Re$ is any function for which $g((x_{j}')_{j\in N_i'})\geq
g((x_{j})_{j\in N_i}$, whenever
$N_i\subset N_i'$, and  $x_{j}'\geq_X x_{j}$ for each $j\in N_i$,
 and $v(x_i;\theta_i,g)$ satisfies increasing differences in $(x_i;\theta_i,g)$.
\end{itemize}
Clearly 4. nests the other cases.  It should be clear that there are many formulations of how an agent's utility depends upon the actions of the neighbors that satisfy increasing differences, as long as the actions of the neighbors enter in a way that interacts monotonically with the agent's own action.

These conditions all involve weak inequalities.
To derive strict implications, we need to introduce a technical condition.
Define $u_i$ to have a smooth dimension if $X= X_{1}\times X_{2}$ in which $X_{1}$ is a compact interval of $\Re$ and $X_{2}$ is a complete lattice,
$u_i$ is continuously differentiable in $x_{1}$, and $\partial u_i/ \partial x_{1}$ is strictly increasing in $\theta_i,d_i$.\footnote{See \citet{vanzandtv2007}.}

Agents choose actions as a function of their types $\theta_i,d_i$.
Let there be some given measure on types $\theta$ in the population denoted $\mu$.
Let agents view their neighbors' degrees as independent across neighbors and independent of the types.
Given are distributions on the degrees of agents other than $i$ in the population $P_i$, and associated $\widetilde{P_i}$ defined by (\ref{tilde}),
which could be functions of $(\theta_i,d_i)$.
Let  $x_i^{friend}(\theta_i, d_i)$ and  $x_i^{soc}(\theta_i, d_i)$ denote Bayesian equilibrium strategies corresponding to beliefs
over neighbors' types and degrees defined by $\mu\times \widetilde{P_i}$  and $\mu\times{P_i}$, respectively.

\begin{proposition}
\label{general}[Network Distortions on Behavior: General Games with Strategic Complements] \   \
Consider a game for which $u_i$ is continuous, bounded, and supermodular in $x_i$, and satisfies increasing differences in $( x_i ; (x_{j})_{j\in N_i} ,\theta_i, d_i) $, for each $i$.
Let ${P}_i$ have weight on at least two degrees and $P_i$ and $\widetilde{P}_i$ be monotone functions of $\theta_i,d_i$.\footnote{Thus, if $\theta_i',d_i'\geq \theta_i,d_i$, then $P_i'\geq P_i$ and $\widetilde{P}_i'\geq \widetilde{P}_i$ in the sense of first-order stochastic dominance.}
Then maximal Bayesian equilibria, $x_i^{friend}$ and $x_i^{soc}$ exist and are nondecreasing in $\theta_i,d_i$.
Moreover,
$$x_i^{friend}(\theta_i, d_i) \geq_i x_i^{soc} (\theta_i,d_i) {\it \ \ for \ all \ }i {\it \ and  \ }\theta_i,d_i.$$
If for each $i$, $u_i$ has a smooth dimension on which $x_{i1}^{friend}(\theta_i, d_i)$ and $x^{soc}_{i1} (\theta_i,d_i)$ are interior for all $(\theta_i,d_i)$, then the inequality is strict for all $\theta_i,d_i$.
\end{proposition}

We also have an immediate corollary that if local externalities are positive (so that $u_i$ is increasing in $(x_{j})_{j\in N_i}$), then the expected utility of the equilibria are ordered in the same way as the actions, while if local externalities are negative ($u_i$ is decreasing in $(x_{j})_{j\in N_i}$) then the welfare ordering is the reverse of the actions.  Incorporating global externalities then requires a comparison between local and global effects, which do not change the results if they move in the same direction, but may lead to ambiguous effects if they conflict in direction, and then the statement requires a large enough negative global externality to reverse the welfare ordering.

The results surrounding the endogenous network also extend, as under appropriate monotonicity conditions higher types prefer higher degrees and more activity.
It is important to note that the endogenous network game is not supermodular.  If a lower preference type increases its degree, then that becomes relatively more frequent as a neighbor, and can partly crowd out  higher types being in a neighborhood in expectation, and so can lower the expected actions of a neighbor.
This does not overturn the logic of the analysis - as degree will still be a non-decreasing function of type and so is behavior.  It just means that the techniques from supermodular games cannot be used in the proofs, and one needs to argue directly based on the monotonicity of strategy choices (and existence comes from continuity and compactness).  Such direct arguments are used in the proof of Proposition \ref{endogenous}, as even the linear-quadratic setting is not supermodular when endogenizing degree.  So one can simply mimic the logic of that proof to extend the endogenous degree choice to more general utility formulations.

It is worth noting that the results here also extend to other interpretations of $d_i$ - beyond it being a degree:  it could also be interpreted as an experience level, or memberships in a number of clubs, and so forth.

\section{Concluding Remarks}

\subsection{Summary}

`Popular' individuals disproportionately impact the perceptions  in a society.
If popular individuals tended to act the same as others this would not systematically bias people's perceptions of others' typical behaviors.
However, as we have shown, there are two ways in which popular individuals and their behaviors differ from others.  First, they have more interactions and that leads them to engage in more activity when facing strategic complementarities.   Second, people who are more predisposed to like a certain behavior will also seek to have more interactions involving that behavior, amplifying the effect.
As shown in this paper,  these two distortions both lead to increases in perceptions of behavior and ultimately feed back to increase the overall behavior in a society.

Depending on the nature of the externalities of an activity the effect of the friendship paradox can be good or bad.  For instance,  these results help us to understand students' systematic over-estimation of their peers' delinquencies that involve social interaction.   Thus, this offers a possible explanation as to why drug and alcohol problems are pervasive in high school and college environments.
Interestingly, the friendship paradox can also be Pareto improving in settings with positive externalities.
It is worth noting that these distortions can be even further exacerbated by social media, where distortions in the number of interactions can be extreme and in which what is posted or communicated is also biased towards behaviors that are social in nature.

\subsection{Empirical Questions and Policy Implications}\label{policy}

The analysis in this paper is theoretical, and it provides a set of clear hypotheses for future empirical work.
That people experience the sorts of complementarities and peer effects modeled here
has been established in a variety of settings (e.g., see \cite{eppler2011,sacerdote2011,ali2011,moretti2011,patacchiniz2008b,dahllm2014}).
Measuring the size of the distortions described in this paper would be of interest, and they are likely to differ across settings with the strengths of the complementarities.

Understanding the friendship paradox's role in the formation of social norms also has policy implications.\footnote{%
This fact has not been lost on marketers and is also an important driver of identifying most-at-risk individuals, for instance using snowball sampling to identify people most at risk for HIV.  Taking advantage of the visibility of friends can also help in fostering adoption of
new programs (e.g., \citet{kimetal2015}).
}
There are many cases in which the analysis of misperceptions due to the friendship paradox apply: people really care about what `normal' behavior is, but extrapolate from
the sample of what they observe.
This case of misperception sheds light on the importance of role models and information access in improving norms.

In particular, our analysis offers an explanation as to why programs known as `social norm marketing' have been successful.
In such programs,  one simply informs people of the true population behavior.
An early instance of such a program was used by Northern Illinois University (see \citet{hainess1996}).   The study found no improvement due to a
traditional educational intervention in which they taught students about dangers of alcohol and emphasized that it was ok to abstain; but then
when they used a new program of informing students of the actual (reported) rates of binge drinking they found significantly improved perceptions of others' rates of binge drinking and reduced binge drinking overall.\footnote{In the base period 43 percent of the students reported binge drinking.
In that same survey just over 69 percent of the students perceived binge drinking as the `norm'.
At the end of the study of the new program, just over 34 percent of students reported binge drinking and 51 percent perceived it as the norm.}
Since then many other social norm marketing programs have been used and studied, including a study by \citet{dejongetal2006} that involved 18 universities with controls, and reached similar conclusions.
Social norm marketing has been tried in a variety of settings, for instance in improving perceptions of others' behavior and
decreasing the incidence of drinking and driving in a controlled trial in Montana (\citet*{perkinsetal2010}).
Other  variants on such programs that our analysis explains are ones that target the highest consuming students, providing them with information about how their behavior ranks compared to the rest of the population (e.g., see  \citet{Agostinellibm1995}).
Our analysis explains why providing information of actual norms should improve perceptions and norms in any settings with complementarities and overall negative externalities, in which people really care about how their behavior matches with the overall population and not just their friends, but base their perceptions of the norm on their own experiences.

Note that our analysis also provides insight regarding situations in which agents hide behaviors - where we can think of the  action above to be either to
avoid a behavior or hide it.   If agents are worried about reputation, then agents who have more interactions might be more likely to hide that they undertake some behavior.   This leads people to underestimate a behavior, and their lowered perceptions of the norm can lead to more hiding of the behavior.   It would be interesting to explore the implications of such results for perceived norms and openness of behavior, for instance, of homosexuality in some societies - and policies such as `don't ask don't tell'.

\bibliographystyle{ecta}
\bibliography{friendshipparadox}

\section*{Appendix}

\noindent{\bf Proof of Lemma \ref{paradox2}}:
Consider a network $g$.
The average degree of agents in the network is
\[
\frac{\sum_{i} d_i(g)}{n} = \frac{1}{n}\sum_{i<j: g_{ij}=1} 2.
\]
The average degree of neighbors is
\[
\frac{1}{n}\sum_{i: d_i(g)>0} \frac{\sum_{j:g_{ij}=1} d_j(g)}{d_i(g)} = \frac{1}{n}\sum_{i<j: g_{ij}=1} \frac{d_j(g)}{d_i(g)} + \frac{d_i(g)}{d_j(g)}.
\]
Thus, it suffices to show that
$$\frac{d_j(g)}{d_i(g)} + \frac{d_i(g)}{d_j(g)}\geq 2$$
and that the inequality is strict if and only if $d_i(g)\neq d_j(g)$.
Note that
$$\frac{d_j(g)}{d_i(g)} + \frac{d_i(g)}{d_j(g)}- 2 = \frac{(d_j(g) - d_i(g))^2}{d_i(g) d_j(g)}. $$
The right hand side of the above equation is nonnegative, and positive if and only if $d_j(g)\neq d_i(g)$.\eproof

\smallskip

\noindent{\bf Proof of Lemma \ref{eq}:}
From the first-order condition of maximizing expected utility (\ref{eu2}), it follows that
the best response of $i$ as a function of $i$'s type and degree is
\begin{equation}
\label{xp}
x_i(\theta_i,d_i) =\frac{\theta_i }{c}+ \frac{a  d_i \widetilde{\E}_i\left[  x_j \right]}{c},
\end{equation}
where $j$ indexes a generic neighbor.

Taking expectations of both sides of the above expression for $x_i(\theta_i,d_i) $ with respect to  $\widetilde{\E}$  (dropping the subscript, given that $P_i$'s are the same for all $i$) yields
\[
\widetilde{\E}\left[x\right] = \frac{\widetilde{\E}\left[\theta\right] }{c}+ \frac{a  \widetilde{\E}\left[  d \right]  \widetilde{\E}\left[  x \right]}{c}.
\]
Thus,
\[
\widetilde{\E}\left[x\right] = \frac{\widetilde{\E}\left[\theta\right] }{c-a  \widetilde{\E}\left[  d \right]  }.
\]

Substituting the above expression into the solution for $x_i(\theta_i,d_i)$ leads to the following characterization of equilibrium,
\[
x^{friend}(\theta_i,d_i)\ = \ \frac{\theta_i }{c}+ \frac{a  d_i \widetilde{\E}\left[x^{friend}\right]}{c} \ = \ \frac{\theta_i }{c}+ \frac{a d_i  \widetilde{\E}\left[\theta\right] }{c\left(c -  a  \widetilde{\E}\left[  d \right]\right)  } ,
\]
as claimed in the lemma.

The same argument with $\E$ replacing $\widetilde{\E}$ produces the expression for $x^{soc}$.\eproof

\smallskip

\noindent{\bf Proof of Proposition \ref{comparison}:}
Recalling from (\ref{etilde}) that
\[
 \widetilde{\E}\left[  d \right] = \sum_{d>0} d  \frac{ P(d) d}{\E[d]}  =\frac{  \E[d^2]}{\E[d]},
 \]
it follows from Lemma \ref{eq}, and the fact that the expectations over $\theta_j$'s of neighbors is the same as the unconditional expectation, that the equilibrium actions are
 \[
x^{friend}(\theta_i,d_i)=
\frac{\theta_i}{c} + \frac{a  d_i \E\left[ \theta \right]}{c(c- a \frac{  \E[d^2]}{\E[d]})}.
\]
and
 \[
x^{soc}(\theta_i,d_i)=
\frac{\theta_i}{c} + \frac{a  d_i \E\left[ \theta \right]}{c(c- a  \E[d])}.
\]

The first part of the proposition, that $x^{friend}(\theta_i,d_i)> x^{soc}(\theta_i,d_i)$ for all $\theta_i$ and $d_i$, then follows from comparing our expression for $x^{friend}(\theta_i,d_i)$ above to that of $x^{soc}(\theta_i,d_i)$, and noting that the only change is in the denominator with a comparison between $\frac{\E[d^2]}{\E[d]}$ and
$ \E[d]$.
The first claim then follows directly since whenever $P$ has a positive variance, then
$$\frac{\E[d^2]}{\E[d]}= \frac{\Var[d] + \E[d]^2}{\E[d]} = \frac{\Var[d] }{\E[d]} + \E[d]> \E[d].$$
This also then implies that $\E[x^{friend}]  > \E[x^{soc}]$, since these are ordered pointwise.

The fact that $\widetilde{\E}[x^{friend}] >\E[x^{friend}]$ follows from the fact that $\widetilde{P}$ strictly first-order stochastically dominates $P$ and $x^{friend}$ is increasing in $d_i$,
which completes the proof.\eproof

\medskip

The following comparative statics on the equilibrium as we change $c,a,P$ are straightforward variations on results in the literature (e.g., \citet{ballestercz2006,galeottigjvy2010}).
These comparative statics offer helpful insight in the proofs of the main results that follow, as they show how varying the distribution of degrees affects the equilibrium, which is one way of thinking about what happens due to the friendship paradox which changes degrees relative to population averages.

For the comparative statics results, the notation $x^{friend}_{a,c,P}, U^{friend}_{a,c,P}$ tracks the dependence of the equilibrium actions and utility functions on the parameters of the setting.

\begin{lemma}
\label{compstats}
[Comparative Statics]

Compare two settings $(a,c,P,\phi)$ and $(a',c',P',\phi)$.\footnote{$\phi$ is held constant since it does not impact actions, only welfare.}
An increase in local complementarities, a decrease in the cost of action, a first-order stochastic dominance increase in the distribution of neighbors' degrees, or a mean-preserving
spread in the degree distribution, all increase equilibrium actions of every type of agent.
That is, if $a\geq a'$, $c\leq c'$, and for the marginal distributions on degrees either $\widetilde{P}\geq_{FOSD} \widetilde{P'}$ or $P$ is a mean-preserving spread of $P'$,\footnote{$\widetilde{P}\geq_{FOSD} \widetilde{P'}$ indicates first-order stochastic dominance.  Note that this condition applies to the distribution of neighbors' degrees, and is {not} implied from stochastic dominance of $P$ over $P'$ (see footnote 19 in \citet{galeottigjvy2010}).  In contrast, the mean-preserving spread is directly on the underlying degree distributions.} with at least one of the inequalities being strict,
then
$x^{friend}_{a,c,P}(\theta_i,d_i) > x^{friend}_{a',c',P'}(\theta_i,d_i)$ for all $i$ and for every $\theta_i,d_i$.  Correspondingly, if $\phi$ is not too negative (there exists $\overline{\phi}< 0$ such that if $\phi \geq \overline{\phi}$), then  $U^{friend}_{a,c,P}(\theta_i,d_i) > U^{friend}_{a',c',P'}(\theta_i,d_i)$, with the reverse inequality if $\phi$ is negative enough (there exists $\underline{\phi}\leq  \overline{\phi}<0$ such that if $\phi \leq \underline{\phi}$).

Similar comparative statics hold for $x^{soc}$, $U^{soc}$, and $U^{naive}$.\footnote{For the case of $x^{soc}$, the first-order stochastic dominance is directly in terms of $P$ and $P'$ rather than $\widetilde{P}, \widetilde{P'}$.  For $U^{naive}$, the comparative static requires a positive enough $\phi$ for the first comparison, as with $\phi$ near 0 the direct interactions dominate and
have ambiguous comparative statics as others' actions increase which is beneficial, but so does the error in best responses which is harmful.}
\end{lemma}

The comparative statics are intuitive.  Increasing the interaction factor, decreasing the cost of actions, and increasing the spread of degrees in the society, all increase the levels of activity by agents and the feedback effects, as well as the amplification due to the friendship paradox.  With positive externalities this benefits an agent, and with negative enough global externalities this harms an agent.  For moderately negative global externalities the welfare impact is ambiguous as the direct interaction externalities are positive and the global ones are negative.


\noindent{\bf Proof of Lemma \ref{compstats}}:
Recall that
\[
x^{friend}(\theta_i,d_i)=
\frac{\theta_i}{c} + \frac{a  d_i \widetilde{\E}\left[ \theta \right]}{c(c- a \widetilde{\E}[d])}.
\]
This is increasing in $a$, decreasing in $c$ (under the maintained assumptions that on $c> a \widetilde{\E}[d]$).
Note that $ \widetilde{\E}[d]$ is increasing as we take a first-order stochastic dominance shift in $\widetilde{P}$, and
also as we take a mean-preserving spread of $P$ since
$\widetilde{\E}[d]=\frac{  \E[d^2]}{\E[d]}$.
The comparative statics in actions follow directly.

From (\ref{up}) we know that
\[
U^{friend}=\frac{c x^{friend}(\theta_i,d_i)^2}{2}  +  \phi \E_i\left[ \sum_{j\neq i} x^{friend}(\theta_j,d_j)\right].
\]

Given that $x^{friend}(\theta_i,d_i)$ increases with
$a$, decrease with $c$, and
increase as we take a first-order stochastic dominance shift in $\widetilde{P}$, and
also as we take a mean-preserving spread of $P$, it follows that
for a nonnegative $\phi$, we end up with a strict increase in the resulting $U$.
So, overall payoffs have gone up for all types.
Given that this is a strict inequality when $\phi$ is 0, across all types and degrees in a compact set, and utilities are continuous in $\phi$, this also holds for some
negative $\phi$'s, establishing the first welfare comparison of the lemma.

Next, note that the equilibrium actions are independent of $\phi$. Given that $\Theta$ is compact and degrees are bounded by $n-1$, and utility and actions are continuous
in types, there is a maximum gain in
\[
\frac{c x^{friend}(\theta_i,d_i)^2}{2}
\]
due to the change from $a',c',P'$ to $a,c,P$.  Call this $X$ (which we know is positive from above, as it corresponds to $\phi=0$).  There is also a change in $\E_i\left[ \sum_{j\neq i} x^{friend}(\theta_j,d_j)\right]$ which is some $Y>0$.
Then provided $X+\phi Y<0$, then the welfare comparison will be negative.  So, setting $\underline{\phi}< - X/Y$ completes the proof.\eproof

\smallskip

Proposition \ref{welfare2b} is a special case of the following proposition, which includes additional comparisons.

\begin{proposition}
\label{welfare2}
[Strict Pareto Rankings and Misperceptions] \  \
Consider a random network model with a degree distribution that has a positive variance and
for which $\widetilde{\E}[\theta_j] = \E[\theta]$.
If $E(\theta)< 2a E[d]/c$ and global externalities are positive or not too negative (there exists $ \overline{\phi}<0$ such that if $\phi \geq \overline{\phi}$),
then:\footnote{$U^{friend}(\theta_i,d_i)  > U^{naive}(\theta_i,d_i) $ holds without either condition  and  $U^{friend}(\theta_i,d_i)  > U^{soc}(\theta_i,d_i) $ holds without the condition $E(\theta)< 2a E[d]/c$.
$U^{naive}(\theta_i,d_i) > U^{soc}(\theta_i,d_i) $ also holds without the condition $E(\theta)< 2a E[d]/c$ if global externalities are positive enough.  }
$$ U^{friend}(\theta_i,d_i) > U^{naive}(\theta_i,d_i) >
U^{soc}(\theta_i,d_i)$$
for all $\theta_i$ and $d_i$.
If global externalities are negative  enough (there exists $\underline{\phi}< \overline{\phi}$ such that if $\phi \leq \underline{\phi}$),
then:
$$U^{soc}(\theta_i,d_i)   >U^{friend}(\theta_i,d_i)
> U^{naive}(\theta_i,d_i)$$
for all $\theta_i$ and $d_i$.
\end{proposition}

\noindent{\bf Proof of Proposition \ref{welfare2}}:
We first develop expressions for the expected utilities for the various scenarios.

From (\ref{eusoc})
\begin{equation}
\label{up0}
U^{friend}(\theta_i,d_i) =  x^{friend}(\theta_i,d_i) \left[ \theta_i - \frac{c}{2} x^{friend}(\theta_i,d_i) + a d_i \widetilde{\E}[x^{friend}(\theta_j,d_j)]  \right]+ \phi \E\left[\sum_{j\neq i} x^{friend}(\theta_j,d_j)\right].
\end{equation}
Substituting the expression for $x^{friend}$ from (\ref{xp}) into the above, it follows  that
\begin{equation}
\label{up}
U^{friend}(\theta_i,d_i) =  \frac{c x^{friend}(\theta_i,d_i)^2}{2} + \phi \E\left[\sum_{j\neq i} x^{friend}(\theta_j,d_j)\right].
\end{equation}
Parallel calculations show that
\begin{equation}
\label{us0}
U^{soc}(\theta_i,d_i) =  x^{soc}(\theta_i,d_i) \left[ \theta_i - \frac{c}{2} x^{soc}(\theta_i,d_i) + a d_i \E[x^{soc}(\theta_j,d_j)]  \right]+ \phi \E\left[\sum_{j\neq i} x^{soc}(\theta_j,d_j)\right].
\end{equation}
\begin{equation}
\label{us}
U^{soc}(\theta_i,d_i) =  \frac{c x^{soc}(\theta_i,d_i)^2}{2} + \phi \E\left[\sum_{j\neq i} x^{soc}(\theta_j,d_j)\right].
\end{equation}
Next, note that
\begin{equation}
\label{um}
U^{naive}(\theta_i,d_i) =  x^{friend}(\theta_i,d_i) \left[ \theta_i - \frac{c}{2} x^{friend}(\theta_i,d_i) + a d_i E[x^{friend}(\theta_j,d_j)]  \right]+ \phi \E\left[\sum_{j\neq i} x^{friend}(\theta_j,d_j)\right].
\end{equation}

So, first let us compare $U^{friend}(\theta_i,d_i)$ and $U^{naive}(\theta_i,d_i)$.
From (\ref{up0}) and (\ref{um}), straightforward calculations show that
\begin{equation}
\label{up-um}
U^{friend}(\theta_i,d_i)-U^{naive}(\theta_i,d_i) =  x^{friend}(\theta_i,d_i)  a d_i \left[ \widetilde{\E}[x^{friend}(\theta_j,d_j)] -\E[x^{friend}(\theta_j,d_j)]  \right]
\end{equation}
which is always greater than 0 by Proposition \ref{comparison}.

Next, let us compare $U^{naive}(\theta_i,d_i)$ and $U^{soc}(\theta_i,d_i)$.
Straightforward calculations starting from (\ref{us0}) and (\ref{um}), and substituting the expressions from (\ref{eqx}) and (\ref{eqxsoc}),  show that

\medskip
\noindent $U^{naive}(\theta_i,d_i)-U^{soc}(\theta_i,d_i) =$
\[
a d_i \E[\theta] \left[ x^{friend}(\theta_i,d_i)  \left(\frac{a {\E}[  d ]}{c}- \frac{\E[\theta]}{2}\right) +\frac{x^{soc}(\theta_i,d_i)  }{2} \right]
\left[\frac{1 }{\left(c -  a  \widetilde{\E}\left[  d \right]\right) } - \frac{1 }{\left(c -  a  {\E}\left[  d \right]\right) }  \right]
\]
\begin{equation}
\label{um-us}
+ \phi \E\left[\sum_{j\neq i} x^{friend}(\theta_j,d_j)-x^{soc}(\theta_j,d_j)\right].
\end{equation}
Under the conditions that $\E[\theta]< 2a {\E}[  d ]/c$ and that $d_i$ has a positive variance (so that
$1/\left(c -  a  \widetilde{\E}\left[  d \right]\right)  >1/ \left(c -  a  {\E}\left[  d \right]\right) $) , the first expression is strictly positive for all $\theta_i,d_i$.
By Proposition \ref{comparison}, the second expression has the sign of $\phi$.   Thus, the expression is positive when $\phi\geq 0$ for all $\theta_i, d_i$.
Given that utilities are continuous and the set of types and degrees is compact, there is a minimum level of difference that is greater than 0.  Thus, it will also hold for
some negative $\phi$'s, greater than some $\overline{\phi}<0$.
Similarly, there is a
maximum level of
the first expression in (\ref{um-us}) across types and degrees.
Call this $X$ (which corresponds to $\phi=0$).
Then, let $Y=\E\left[\sum_{j\neq i} x^{friend}(\theta_j,d_j)-x^{soc}(\theta_j,d_j)\right]>0$.
Then provided that $\phi$ is low enough so that  $X+\phi Y<0$, then the welfare comparison will be negative, which completes the proof of the ordering of $U^{soc}(\theta_i,d_i)$ and $U^{naive}(\theta_i,d_i)$.

The orderings between $U^{friend}(\theta_i,d_i)$ and $U^{soc}(\theta_i,d_i)$ follow from the proof of Lemma \ref{compstats}, noting that difference the $x^{soc},U^{soc}$ and $x^{friend},U^{friend}$ just corresponds
to a change in the use of $P$ versus $\widetilde{P}$, which is a strict mean-preserving spread.   Note that this comparison does not require the assumption that
$\E[\theta]< 2a {\E}[  d ]/c$. \eproof

\smallskip

To prove Proposition \ref{inequality} we show a more complete proposition,  Proposition \ref{inequality3}.

\begin{proposition}
\label{inequality3}
Consider a random network model $(a,c,\phi, P)$, that has a degree distribution that has a positive variance and
for which $\widetilde{\E}[\theta_j] = \E[\theta]$.
Consider any $i$ and two different degrees $d_i>d_i'$.
Then
\[
x^{friend}(\theta_i,d_i) - x^{friend}(\theta_i,d_i')  > x^{soc}(\theta_i,d_i) - x^{soc}(\theta_i,d_i').
\]
Also, if $E(\theta)< 2a E[d]/c$ then\footnote{The condition is a sufficient condition for $U^{naive}(\theta_i,d_i) - U^{naive}(\theta_i,d_i') > U^{soc}(\theta_i,d_i) - U^{soc}(\theta_i,d_i')$ while $U^{friend}(\theta_i,d_i) - U^{friend}(\theta_i,d_i') > U^{naive}(\theta_i,d_i) - U^{naive}(\theta_i,d_i')$  holds without the condition.}
\[
U^{friend}(\theta_i,d_i) - U^{friend}(\theta_i,d_i') > U^{naive}(\theta_i,d_i) - U^{naive}(\theta_i,d_i') > U^{soc}(\theta_i,d_i) - U^{soc}(\theta_i,d_i')
\]
for all $\theta_i$.
\end{proposition}

\noindent{\bf Proof of Proposition \ref{inequality3}}:
The comparisons between $x^{soc}$ and $x^{friend}$ follow from proof of Proof of Proposition \ref{inequality2}, below, noting that difference the $x^{soc}$ and $x^{friend}$ just corresponds
to a change in the use of $P$ versus $\widetilde{P}$, which is a strict mean-preserving spread.

To compare $U^{naive}(\theta_i,d_i) - U^{naive}(\theta_i,d_i') $ to $ U^{soc}(\theta_i,d_i) - U^{soc}(\theta_i,d_i')$, note that by (\ref{um-us}), $U^{naive}(\theta_i,d_i)-U^{soc}(\theta_i,d_i) $ is increasing in $d_i$, which implies the comparison.
Similarly to compare $U^{friend}(\theta_i,d_i) - U^{friend}(\theta_i,d_i')$ to $ U^{naive}(\theta_i,d_i) - U^{naive}(\theta_i,d_i')$  note that by  (\ref{up-um})
$U^{friend}(\theta_i,d_i)-U^{naive}(\theta_i,d_i) $ is increasing in $d_i$.\eproof

\smallskip

\begin{proposition}
\label{inequality2}
[Increased Inequality, Part II]

Compare two settings $(a,c,P,\phi)$ and $(a',c',P',\phi)$.\footnote{Changes in $\phi$ do not impact actions, only welfare.}
An increase in local complementarities, a decrease in the cost of action, a first-order stochastic dominance increase in the distribution of neighbors' degrees, or a mean-preserving
spread in the degree distribution, all increase equilibrium actions of every type and the equilibrium
utility of every type of agent.
That is, if $a\geq a'$, $c\leq c'$ and either $\widetilde{P}\geq_{FOSD} \widetilde{P'}$ or $P$ is a mean-preserving spread of $P'$, with at least one of the inequalities being strict,
then
\[
x^{friend}_{a,c,P}(\theta_i,d_i) - x^{friend}_{a,c,P}(\theta_i,d_i') > x^{friend}_{a',c',P'}(\theta_i,d_i) - x^{friend}_{a',c',P'}(\theta_i,d_i')
 \]
 and
 \[
U^{friend}_{a,c,P}(\theta_i,d_i) - U^{friend}_{a,c,P}(\theta_i,d_i') > U^{friend}_{a',c',P'}(\theta_i,d_i) - U^{friend}_{a',c',P'}(\theta_i,d_i')
 \]
for all $i$ and $\theta_i$ and $d_i>d_i'$.
\end{proposition}

\smallskip

\noindent{\bf Proof of Proposition \ref{inequality2}}:
To see the first claim, note that
\[
x^{friend}_{a,c,P}(\theta_i,d_i) - x^{friend}_{a,c,P}(\theta_i,d_i') = \frac{a  (d_i - d_i') \E\left[ \theta \right]}{c(c- a \widetilde{\E}[d])}.
 \]
This is increasing in $a$, decreasing in $c$,  and increasing in $\widetilde{\E}[d]= \frac{\E[d^2]}{\E[d]}$ which increases when either $\widetilde{P}\geq_{FOSD} \widetilde{P'}$ or $P$ is a mean-preserving spread of $P'$ (at least one strict).
This establishes the first part of the result.

Note that substituting $x^{friend}(\theta_i,d_i)$ into (\ref{eu2}) it follows that
\[
U^{friend}(\theta_i,d_i) =\frac{\left(\theta_i  + a   d_i \widetilde{\E}_i\left[ x^{friend}(\theta_j,d_j) \right]\right)^2}{2c} +  \phi \E_i\left[ \sum_{j\neq i} x^{friend}(\theta_j,d_j)\right].
\]
Thus,
\[
U^{friend}(\theta_i,d_i) - U^{friend} (\theta_i,d_i') = \frac{\left(\theta_i  + a   d_i \widetilde{\E}_i\left[ x^{friend}(\theta_j,d_j) \right]\right)^2
- \left(\theta_i  + a   d_i' \widetilde{\E}_i\left[ x^{friend}(\theta_j,d_j) \right]\right)^2}{2c}
\]
Notice that this expression is increasing in $a$ and $\widetilde{\E}_i\left[ x^{friend}(\theta_j,d_j)\right]$ and decreasing in $c$.
Given that $\widetilde{\E}_i\left[ x^{friend}(\theta_j,d_j)\right]$ increases as we make the claimed changes from $a',c',P'$ to $a,c,P$, the result then follows.\eproof

\smallskip

\noindent{\bf Proof of Proposition \ref{endogenous}}:
Consider a pure strategy symmetric equilibrium in which agents put positive probability on a positive degree - so that there is some interaction.
Let us first analyze the friend game and this equilibrium.

First note that in any such symmetric equilibrium, it must be that
since agents are replying when choosing degrees and action levels, their action levels must also be best replies to the other action levels holding their degree choices fixed. Thus, taking the first-order conditions of (\ref{eud}) with respect to $x_i$ the equilibrium $x_i$'s satisfy
\begin{equation}
\label{xendog}
x^{friend}_{end}(\theta_i,d(\theta_i))=  \frac{\theta_i }{c}+ \frac{a  d(\theta_i) \widetilde{\E}\left[  x^{friend}_{end}(\theta_j,d(\theta_j)) \right]}{c}
=
\frac{\theta_i}{c} + \frac{a  d(\theta_i) \widetilde{\E}\left[ \theta_j \right]}{c(c- a \widetilde{\E}[d(\theta_j)])}.
\end{equation}
The second equality follows from solving for the equilibrium values as in Lemma \ref{eq}, taking the $d(\theta)$ choices as given, with the
only difference being that now the numerator has the expectation $\widetilde{\E}\left[ \theta_j \right]$ (rather than ${\E}\left[ \theta_j \right]$), which conditions on the fact that the types of neighbors now correlates with their degrees.

Next, consider some $i$ and let us examine the best response choices of $d_i$, knowing that these must be best responses anticipating that actions will be according to $x^{friend}_{end}$.

Let us first consider the choice of an agent as if he or she were maximizing a continuous random variable.
Taking the first-order conditions of (\ref{eud}) with respect to $d_i$ (and invoking the Envelope Theorem with respect to $x^{friend}_{end}(\theta_i,d_i)$ as a function of $d_i$) for the maximization of $i$'s expected utility leads to
\begin{equation}
\label{endd}
 a   x^{friend}_{end}(\theta_i,d_i)  \widetilde{\E}\left[  x^{friend}_{end}(\theta_j,d_j(\theta_j)) \right]- k d_i =0.
\end{equation}

The second derivative of the expected utility is
\[
\frac{\partial x^{friend}_{end}(\theta_i,d_i)}{\partial d_i} a \widetilde{\E}\left[  x^{friend}_{end}(\theta_j,d_j(\theta_j)) \right]- k .
\]
From (\ref{xendog}) it follows that
$
\frac{\partial x^{friend}_{end}(\theta_i,d_i)}{\partial d_i}= \frac{a \widetilde{\E}\left[x^{friend}_{end}(\theta_j,d_j(\theta_j)) \right] }{c  }
$
and so the second derivative is
\[
\frac{a^2 \widetilde{\E}\left[x^{friend}_{end}(\theta_j,d_j(\theta_j)) \right]^2 }{c  } - k ,
\]
which is negative by assumption (Footnote \ref{costs} implies that $ck> a^2(n-1)^2$, which is
at least $ a^2 \widetilde{\E}\left[x^{friend}_{end}(\theta_j,d_j(\theta_j)) \right]^2$)\footnote{Note that by taking 
 expectations of both sides of (\ref{endd}), it follows that $ a \widetilde{\E}\left[x^{friend}_{end}(\theta_j,d_j(\theta_j)) \right]^2 = k \widetilde{\E}\left[d_j(\theta_j) \right], $ and the right hand side is no more than $k (n-1)\leq a (n-1)^2$ .} .

Thus, the expected utility is strictly concave in $d_i$ and has a maximum at a point solving
\[
kd_i= a   x^{friend}_{end}(\theta_i,d_i)  \widetilde{\E}\left[  x^{friend}_{end}(\theta_j,d_j(\theta_j)) \right].
\]
Then, substituting for $x^{friend}_{end}(\theta_i,d_i)$ and solving for $d_i$, the optimal degree ignoring integer constraints is:
\[
d_i^* = \frac{\theta_i}{c} \left[ \frac{ca \widetilde{\E}\left[x^{friend}_{end}(\theta_j,d_j(\theta_j))\right] }{ck - a^2 \widetilde{\E}\left[  x^{friend}_{end}(\theta_j,d_j(\theta_j)) \right]^2 } \right].
\]

Given the strict concavity of the expected utility function, the maximizing integer choice for $d^{end}$ must put probability only on either highest integer that does not exceed $d_i^*$ or the lowest
one that is not smaller than $d_i^*$.

Next, note that (from (\ref{eud}))
\[
\frac{\partial^2 EU_i(\theta_i,d_i)}{\partial \theta_i \partial d_i} = a  \frac{\partial x^{friend}_{end}(\theta_i,d_i)}{\partial \theta_i}  \widetilde{\E}\left[  x^{friend}_{end}(\theta_j,d_j(\theta_j)) \right]  =  a   \widetilde{\E}\left[  x^{friend}_{end}(\theta_j,d_j(\theta_j)) \right]/c >0 .
\]
This implies, together with the strict concavity of utility in $d_i$, that if some $\theta_i$ weakly prefers $d_i$ to some lower $d_i'$, then any higher type strictly prefers the higher degree.
This implies that the the optimal $d^{end}(\cdot)$ is nondecreasing, and that at most a set of measure 0 of types will be indifferent between two degrees, and
so the strategy can be taken to be pure.

The comparison between $x^{friend}_{\perp}(\theta_i,d^{end}(\theta_i))$ and $ x^{soc}_{\perp}(\theta_i,d^{end}(\theta_i))$ follows from Proposition \ref{comparison}, just substituting in the induced equilibrium degree distribution  (and $ x^{soc}_{\perp}=  x^{soc}_{end}$).

To make the comparison between
$x^{friend}_{end}(\theta_i,d^{end}(\theta_i))$ and $ x^{friend}_{\perp}(\theta_i,d^{end}(\theta_i))$, note that the only difference is that
\[
x^{friend}_{end,i}=
\frac{\theta_i}{c} + \frac{a  d_i \widetilde{\E}\left[ \theta \right]}{c(c- a \frac{  \E[d^2]}{\E[d]})}
\]
while
  \[
x^{friend}_{\perp, i}=
\frac{\theta_i}{c} + \frac{a  d_i {\E}\left[ \theta \right]}{c(c- a \frac{  \E[d^2]}{\E[d]})},
\]
and so it boils down to a comparison between $\widetilde{\E}\left[ \theta \right]$ and ${\E}\left[ \theta\right]$.
Note that
\[
 \widetilde{\E}\left[  \theta \right] = \sum_d  \E[\theta | d^{end}(\theta) =d]  \frac{ P(d) d}{\E[d]}  >   \sum_d  \E[\theta | d^{end}(\theta) =d ] P(d)  =\E\left[  \theta \right],
 \]
 whenever there are at least two degrees chosen in equilibrium.  This follows from the fact that $d^{end}(\theta)$ is nondecreasing in $\theta$ and pure, and so $\E[\theta | d^{end}(\theta) =d]$ is increasing in $d$, together with the fact that $\frac{ P(d) d}{\E[d]} $ strictly first-order stochastically dominates $P(d)$,
 which completes the proof.\eproof

 \smallskip

\noindent{\bf Proof of Proposition \ref{general}}:
 The existence and monotonicity of greatest equilibria, in both the network and population matching cases, follows from Proposition 14 in \citet{vanzandtv2007}.
The ordering between actions follows from the fact that $\widetilde{P}_i$ strictly first-order stochastically dominates $P_i$ (given that $P_i$ has weight on at least two degrees)
and Proposition 16 in \citet{vanzandtv2007}, as we can view the only difference between $x^{friend}_i$ and $x^{soc}_i$ as a change in the distributions over neighbors' degrees.  The distribution over other agents' types is unchanged, given the independence, and then to see the first-order stochastic dominance, note that $\widetilde{P}_i(d)/P_i(d) = d/\E[d]$ which is strictly increasing in $d$.
The strict ordering of actions in the case with a smooth dimension and all interior actions then follows from their Corollary 17.\footnote{Note that their proof extends to the case in which the stochastic dominance is strict only on one dimension of agents' types, here degrees, and that dimension drives a strict increase in the derivative of utility with
respect to the smooth dimension of actions.}\eproof

\section*{Public Goods and Strategic Substitutes}

The results above concern games of strategic complements.  That is a case of fundamental interest since many interactions fall into that category.
Games with strategic substitutes also apply to many settings, such as those in which agents share tasks or contribute to local public goods.
Let me briefly discuss how the results change in the case of substitutes.

With strategic substitutes the interaction of incentives between agents is reversed compared that under  strategic complements.
In a game of local strategic substitutes,
 $i$'s utility is again described by a function of the form
\[
u_i(x_i, (x_{j})_{j\in N_i}, \theta_i, d_i),
\]
in which we maintain the same assumptions as in the case of complementarities before, {\sl except} that we reverse the direction of how $ (x_{j})_{j\in N_i}$ and $d_i$
affect changes in utility with regards to changes in $x_i$. In particular, in this case $u_i$ satisfies increasing differences in $( x_i ; (-x_{j})_{j\in N_i} ,\theta_i, -d_i) $.\footnote{The sign on $\theta_i$ is not reversed,
as this still captures an agent's personal predisposition for the behavior.}
So, agents prefer to take lower actions if they have more neighbors and/or those neighbors take higher actions.  It is still possible that agents have utility that increases in  $ (x_{j})_{j\in N_i}$ and $d_i$, but the incentives to choose a higher $x_i$ decreases as an agent sees more activity by others in their neighborhood.  This applies to standard local public goods games.

In such a setting, using similar arguments to those behind the results above, with a sign reversal, it follows that
$x^*(\theta_i,d_i)$ is nondecreasing in $\theta_i$ and {\sl nonincreasing} in $d_i$.
This then also provides implications for the friendship paradox.
When matched with higher degree neighbors (presuming independence between $\theta$s and $d$s),
one expects less activity from those neighbors than when matched with lower degree neighbors.
Thus, the network setting leads agents to expect less action by their neighbors than in the benchmark population matching setting, and so this ultimately leads
agents to {\sl increase} their own actions in response.

To see how this works in more detail, let us consider a canonical example.   The example is that of a best-shot public goods game (e.g., see \citet{galeottigjvy2010}).

In this setting, each agent chooses an action $x_i\in\{0,1\}$ - whether to provide a local public good.
Providing the good costs $c>0$.   The agent's payoff is then the max of the actions in his or her neighborhood, including her own action.
In particular, the payoff is
\[
\theta_i I_{\left[ x_i +  \sum_{j\in N_i} x_j>0\right]} - c x_i,
\]
where $I$ is the indicator function.
This applies to settings in which if one agent invests in the public good then all of her friends can share in the value of the good.  Examples include completing a task, or buying a book that can be lent to friends, or acquiring information that can be shared with the friends (but for simplicity does not transfer multiple hops).
Each agent would prefer that a neighbor provide the good, but would rather provide the good if no neighbor does.

Let $\widetilde{\pi}_i$ denote the probability agent $i$ perceives that any given one of her neighbors will provide the public good.
Then agent $i$ prefers providing the public good if
\[
\theta_i - c \geq \theta_i [1-(1-\widetilde{\pi}_i)^{d_i}]
\]
or (presuming that $\widetilde{\pi}_i<1$)
\[
\theta_i \geq \frac{c}{(1-\widetilde{\pi}_i)^{d_i}}.
\]

Thus, presuming that $\widetilde{\pi}_i<1$, there is a threshold
$$t_i(d_i)=\frac{c}{(1-\widetilde{\pi}_i)^{d_i}} >0$$
for which the agent's best response is to provide the public good ($x_i=1$) if $\theta_i> t_i(d_i)$ and not to provide the public good ($x_i=0$) if $\theta_i< t_i(d_i)$.\footnote{In a case in which $\theta_i$ has an atomless distribution, the indifferent case is negligible and otherwise there may be some mixing at the precise threshold of $t_i(d_i)$.}
Note that $t_i(d_i)$ is increasing
in $d_i$.

When the distribution of neighbors' degrees and types is the same across agents, the probability that a random neighbor will provide the public good in a
symmetric equilibrium, denoted by $\widetilde{\pi}$,
is then
\[
\widetilde{\pi} = \sum_{d} Pr\left[ \theta > t(d) \right]  \widetilde{P}(d).
\]
In equilibrium this must solve
\[
\widetilde{\pi} =  \sum_{d} Pr\left[ \theta > \frac{c}{(1-\widetilde{\pi})^{d}} \right]  \widetilde{P}(d).
\]
Given that the right hand side is decreasing in $\widetilde{\pi}$ and is positive when $\widetilde{\pi}=0$ (presuming that $c$ lies in the support of $\theta$), this has a unique solution, associated with the unique symmetric equilibrium.

Next, note that first-order stochastic dominance shifts in $\widetilde{P}(d)$ lead the right hand side to decrease for every value of $\widetilde{\pi}$ and so the equilibrium value of $\widetilde{\pi}$ must
decrease.\footnote{Higher values of $d$ lead to higher values of $\frac{c}{(1-\widetilde{\pi})^{d}}$, which lead to lower values of $Pr\left[ \theta_j > \frac{c}{(1-\widetilde{\pi})^{d}} \right]$.  }
This in turn, leads to a lower value of $t(d)$, since it is increasing in $\widetilde{\pi}$.

Thus, it follows that $t^{soc}(d_i) > t^{friend}(d_i)$ for every $d_i$, and so the thresholds are lower under the friendship paradox.  This means that there is more public good provision by
all degrees of agents under the friendship paradox, but this comes from the reaction to an overall lower expected probability that a random neighbor provides the public good under the friendship paradox.

Note, however, that the expected utility of an agent of any given degree generally tends to go down in the network setting compared to the random matching setting, since agents are matched with agents of higher degrees and expect less activity from their neighbors overall.
Thus, even though the network setting incentivizes more activity by agents, this is because they are more frequently matched with high degree agents who tend to free-ride more on the action.  This leads to lower expected utilities by each type of agent and overall.

In games of strategic substitutes, endogenizing the network leads to ambiguous effects on overall actions and welfare.
In most such settings, people who have higher payoffs from the activity also tend to benefit from having higher degree (presuming that there is some marginal gain from
neighbors' provision of the public good on top of one's own provision).\footnote{In the case of the best-shot public goods
game, the endogenous network formation game becomes degenerate.  Any agent who intends to provide the public good in that game gets no additional value from having neighbors.
 Thus, the only agents who would choose to pay to have connections would be those planning not to provide the public good - but then they would not want to have connections in that case.
 To have a nontrivial game in which anyone forms connections, agents have to be endowed with some base degree.  In that case, in equilibrium, only the lowest degree agents would provide the public good.  Those agents actually turn out to be the higher $\theta$ agents in this particular game.}  This leads to an overall ambiguous effect as
agents' high type pushes them to take higher actions but their increased endogenous degree tends to reduce their actions - and the overall effect depends on the parametric specification.
Thus, while the results from the strategic complements in terms of comparisons on a fixed network have (reversed) analogs in the case of strategic substitutes, the case in which the network is endogenized does not extend.
This means that the overall impact of the friendship paradox in the case of strategic substitutes can be ambiguous and will depend on details of the preferences - whether the individual incentives to provide the good or the local externalities dominate.

Although I have analyzed the case of the best-shot public goods game, the reasoning extends to more general games, similarly to the way that the linear-quadratic results generalized in the previous section.

\end{document}